\documentclass{article}
\usepackage[utf8]{inputenc}
\usepackage{amsmath,amsthm,amsfonts}
\usepackage[margin=1in]{geometry}
\usepackage{mathrsfs}
\usepackage{graphicx}
\usepackage[linktocpage,linkcolor=red]{hyperref}
\hypersetup{colorlinks, allcolors=red}
\usepackage{makecell}
\usepackage{newtxmath}
\usepackage{thm-restate}
\usepackage{csquotes}
\usepackage[capitalise]{cleveref}
\usepackage{algpseudocode}
\allowdisplaybreaks

\usepackage{fancybox, pdfsync}
\newtheorem{definition}{Definition}
\usepackage[noend,linesnumbered]{algorithm2e}

\newenvironment{fminipage}
  {\begin{Sbox}\begin{minipage}}{\end{minipage}\end{Sbox}\fbox{\TheSbox}}

\usepackage{bbm}
\newtheorem{theorem}{Theorem}[section]
\newtheorem{corollary}[theorem]{Corollary}

\newtheorem{lemma}[theorem]{Lemma}

\newtheorem{fact}[theorem]{Fact}

\newtheorem{remark}[theorem]{Remark}
\newtheorem*{remark*}{Remark}

\usepackage{float}
\usepackage{color}

\def\abs#1{\left|#1  \right|}

\def\norm#1{\left\| #1 \right\|}

\newcommand\MM{\boldsymbol{\mathit{M}}}

\newcommand\F{\mathbb{F}}
\newcommand\R{\mathbb{R}}
\newcommand\Z{\mathbb{Z}}

\newcommand\N{\mathbb{N}}

\usepackage{pgffor}
\foreach \x in {A,...,Z}{
	\expandafter\xdef\csname m\x\endcsname{\noexpand\mathbf{\x}}
}
\foreach \x in {A,...,Z}{
	\expandafter\xdef\csname om\x\endcsname{\noexpand\overline{\noexpand\mathbf{\x}}}
}
\foreach \x in {A,...,Z}{
	\expandafter\xdef\csname c\x\endcsname{\noexpand\mathcal{\x}}
}

\renewcommand{\th}{^{th}}

\renewcommand{\tilde}{\widetilde}
\renewcommand{\hat}{\widehat}

\newcommand\Otil{\tilde{O}}

\newcommand\update{\textsc{Update}}

\newcommand{\vertiii}[1]{{\left\vert\kern-0.25ex\left\vert\kern-0.25ex\left\vert #1 
    \right\vert\kern-0.25ex\right\vert\kern-0.25ex\right\vert}}

\newcommand{\fro}{\textnormal{F}}
\newcommand{\veps}{\varepsilon}
\DeclareMathOperator*{\poly}{poly}
\DeclareMathOperator*{\rank}{rank}

\newcommand{\vb}{\mathbf{b}}

\newcommand{\vu}{\mathbf{u}}

\newcommand{\vecv}{\mathbf{v}}

\newcommand{\vx}{\mathbf{x}}

\newcommand{\ma}{\mathbf{A}}

\newcommand{\mb}{\mathbf{B}}

\newcommand{\mc}{\mathbf{C}}
\newcommand{\md}{\mathbf{D}}

\newcommand{\mi}{\mathbf{I}}

\newcommand{\ml}{\mathbf{L}}
\newcommand{\mm}{\mathbf{M}}

\newcommand{\mn}{\mathbf{N}}

\newcommand{\mq}{\mathbf{Q}}
\newcommand{\mr}{\mathbf{R}}
\newcommand{\ms}{\mathbf{S}}

\newcommand{\mv}{\mathbf{V}}
\newcommand{\matu}{\mathbf{U}}
\newcommand{\mw}{\mathbf{W}}

\newcommand{\mx}{\mathbf{X}}

\newcommand{\my}{\mathbf{Y}}

\newcommand{\mz}{\mathbf{Z}}

\newcommand{\mzero}{\mathbf{0}}

\DeclareMathOperator{\tr}{tr}

\newcommand{\eeEntryExp}{1.405}
\newcommand{\neEntryUpdateExp}{1.528}
\newcommand{\neEntryQueryExp}{0.528}
\newcommand{\neEntryUpdateExpPlusOne}{2.528}

\usepackage{enumitem}
\usepackage{authblk}

\makeatletter
\def\@fnsymbol#1{\ensuremath{\ifcase#1\or \dagger\or \ddagger\or
   \mathsection\or \mathparagraph\or \|\or **\or \dagger\dagger
   \or \ddagger\ddagger \else\@ctrerr\fi}}
    \makeatother

\begin{document}

\title{
The Bit Complexity of Dynamic Algebraic Formulas\\
and their Determinants
}
\date{}

\author[]{Emile Anand}
\affil[1]{California Institute of Technology}
\author[2]{Jan van den Brand}
\author[3]{Mehrdad Ghadiri}
\author[4]{Daniel Zhang}
\affil[2,4]{Georgia Institute of Technology}
\affil[3]{Massachusetts Institute of Technology}

\maketitle

\begin{abstract}
    Many iterative algorithms in optimization, computational geometry, computer algebra, and other areas of computer science require repeated computation of some algebraic expression whose input changes slightly from one iteration to the next.
Although efficient data structures have been proposed for maintaining the solution of such algebraic expressions under low-rank updates, most of these results are only analyzed under exact arithmetic (real-RAM model and finite fields) which may not accurately reflect the complexity guarantees of real computers. In this paper, we analyze the stability and bit complexity of such data structures for expressions that involve the inversion, multiplication, addition, and subtraction of matrices under the word-RAM model. 
We show that the bit complexity only increases linearly in the number of matrix operations in the expression. In addition, we consider the bit complexity of maintaining the determinant of a matrix expression. We show that the required bit complexity depends on the logarithm of the condition number of matrices instead of the logarithm of their determinant. We also discuss rank maintenance and its connections to determinant maintenance. Our results have wide applications ranging from computational geometry (e.g., computing the volume of a polytope) to optimization (e.g., solving linear programs using the simplex algorithm).
\end{abstract}

\newpage

\tableofcontents

\thispagestyle{empty}
\newpage
\pagenumbering{arabic}
\section{Introduction}

Computing algebraic expressions is a workhorse of many iterative algorithms in modern optimization, computational geometry, and dynamic algorithms. Examples include but are not limited to interior point methods for solving linear programs ~\cite{lee2013path,CLS21,jiang2020fasterb,Brand20,BrandLSS20}, iterative refinement for solving $p$-norm regression problems \cite{bubeck2018homotopy,AdilKPS19,AdilPS19,AdilS20,jiang2022dynamic}, semi-definite programming \cite{huang2022solving,jiang2020faster,lin2023online}, and many algorithmic graph theory problems \cite{Sankowski04,van2019dynamicb,BergamaschiHGWW21,chen2022maximum}.

Such algebraic expressions are usually represented as matrix formulas involving matrices and operations such as inversion, multiplication, and addition/subtraction. In many iterative algorithms, the algebraic expression does not change over the course of the algorithm, and only low-rank updates occur to the corresponding matrices from one iteration to the next. For example, for $\ma,\mb\in\R^{n\times n}$, if we have $\ma\mb$ from a previous iteration and one column of $\mb$ changes in the next iteration, we can update $\ma \mb$ in $O(n^2)$ time which is much faster than computing $\ma \mb$ from scratch again.
This has been exploited in many iterative algorithms to reduce the amortized cost of iteration and, therefore, the total running time of the algorithms.

A main component of this approach is the Sherman-Morrison-Woodbury (SMW) identity (see \eqref{eq:woodbury}), which informally states that the inverse of a rank-$k$ perturbation of a matrix $\ma$ can be obtained by a rank-$k$ perturbation of $\ma^{-1}$. Although this identity (also called \emph{inverse maintenance}) has been used from the early days of optimization and control theory, it was recently shown that \emph{any matrix formula} involving only inversion, multiplication, and addition/subtraction operations can be maintained under low-rank updates with the SMW identity \cite{Brand21}. 
The main idea is to inductively construct a large matrix whose inverse contains a block that is precisely the output of the formula. 

The result of \cite{Brand21} is under the \emph{real-RAM model}, which assumes each arithmetic operation is carried over to infinite precision in constant time. Although this is a valid assumption for finite fields, it does not hold over real numbers. For example, in modern computers, floating-points are the number system of choice that only has a finite precision. Then, it is unclear whether such inverse maintenance techniques are sufficiently stable so that the downstream iterative algorithm outputs the correct solution (e.g., whether the iterative optimization algorithm converges).

Very recently, \cite{MehrdadPV23} analyzed the SMW identity over fixed-point arithmetic and showed that a bit complexity proportional to the logarithm of the condition number (ratio of the largest singular value to the smallest singular value) of the corresponding matrix is sufficient to guarantee the stability of the inverse maintenance over the \emph{word-RAM model} in which the running time of arithmetic operations is proportional to the number of bits of corresponding numbers and only finite precision is guaranteed (and the precision itself depends on the number of utilized bits).

This implies that in order to show that the techniques of \cite{Brand21} also hold over the word-RAM model, we need to bound the condition number of the inductively constructed matrix for arbitrary matrix formulas. Indeed, we affirmatively show that the condition number of the constructed matrix is $\kappa^{O(s)}$, where $\kappa$ is an upper bound for the condition numbers of the input matrices and $s$ is the number of input matrices. This implies that a bit complexity of $O(s\log \kappa)$ is sufficient to guarantee the stability of dynamically updating the matrix formulas. We point out that a naive analysis would give a bound of $O(2^s \log \kappa)$ for the bit complexity, and our bound on the condition number is asymptotically tight since one can easily see that the product of $s$ matrices each with condition number $\kappa$ results in a matrix with condition number $\kappa^s$, e.g., consider $\ma^s$.

In addition, we consider the stability and bit complexity of maintaining the determinant and rank of matrix formulas with inversion, multiplication, and addition/subtraction. An application of maintaining the determinant is in the faster computation of the volume of a polytope \cite{FisikopoulosP16}, and an application of the rank maintenance is in dynamic maximum matching \cite{10.5555/1283383.1283397}.

To maintain the determinant of a matrix formula up to a multiplicative error of $(1\pm \epsilon)$ for $0<\epsilon<1$, in addition to the inductively constructed matrix $\mn$ of \cite{Brand21}, we construct another matrix $\widehat{\mn}$, and show that the determinant of the matrix formula is the ratio of $\det(\widehat{\mn})$ to $\det(\mn)$. This then allows us to use the \emph{matrix determinant lemma} (see \eqref{eq:determinantlemma}) to maintain the determinant. Although one might expect that we would require $\log\det(\mn)$ number of bits for determinant maintenance, we show that $O(s \log (\kappa/\epsilon))$ bits are sufficient. Note that $\log \kappa$ is preferable to $\log \det$ since, for random matrices, the condition number is polynomial in the dimension of the matrix with high probability \cite{edelman1988eigenvalues,edelman1989eigenvalues} while the determinant is exponential \cite{tao2005random}.

We also consider rank maintenance over finite fields. This is because, under fixed-point arithmetic, we can multiply our matrices by a large number to obtain integer matrices and then perform all operations modulo a sufficiently large prime number ($\poly(n)$ is sufficient). Then the rank of such matrix formula over $\Z_p$ is the same as the rank of the original matrix formula with high probability.

We believe optimizers and algorithm designers can use our results as black boxes to analyze their algorithms under the word-RAM model. The only additional part on their side is to analyze what error bounds can be tolerated in the corresponding algorithm while guaranteeing the returned outputs are correct. Then, our results provide the corresponding running time and bit complexity bounds for the required errors.

Our algorithmic results are presented as dynamic data structures in the next \cref{sec:our_result}. They cover the most common update schemes occurring in iterative algorithms, such as updating one entry of the matrix and querying one entry or updating a column and querying a row. 

Finally, to illustrate the effectiveness of our approach and results, we discuss two example applications. The first one, discussed in \cref{subsec:basic_solution}, considers finding a basic solution of a set of linear constraints in the standard form $\ma \vx = \vb$, $\vx \geq 0$ where $\ma\in\R^{d\times n}$ and $n\geq d$. The operations involved in this algorithm are similar to the simplex algorithm. Beling and Megiddo \cite{Beling_Megiddo_1998} presented two algorithms, a simple one with $O(d^2 n)=O(n^3)$ time, and a more complicated one with $O(d^{\neEntryUpdateExp}n)=O(n^{2.528})$ time. 
Both these algorithms assumed the real-RAM model ($O(1)$ time per arithmetic operation with infinite precision).
We show that for $n=O(d)$, by simply plugging our data structures into the simple algorithm, the time complexity becomes 
$\Otil(n^{2.528} \log(\kappa \cdot \max\det))$ in bit complexity (i.e., number of bit operations).
Here $\max\det$ and $\kappa$ is the maximum determinant and condition number over all square $d\times d$ submatrices. 
In particular, for matrices with $\log\max\det = \poly\log(n)$ (as is the case when modelling many combinatorial problems as linear programs), our worst-case running time (i.e., number of bit operations) is $\Otil(n^{2.528} \log \kappa)$.
Thus, not only does the simple algorithm become competitive with the more complicated algorithm, we also show that it can be efficiently implemented without the real-RAM assumptions.

The second example application is for dynamically maintaining the size of the maximum matching of a graph that goes through edge deletion/insertion. 
We show in \Cref{subsec:dynamic_max_matching} that our rank maintenance data structure can be used for this purpose with a cost of $O(n^{\eeEntryExp})$ arithmetic operations per update.

\subsection{Our Results}
\label{sec:our_result}

Our first result is the following generic data structure that can maintain the value of any matrix formula.
Here a matrix formula is any expression that can be written using the basic matrix operations of addition, subtraction, multiplication, and inversion.

\begin{restatable}{theorem}{thmFormulaMaintenance}
    \label{application: general datastructures theorem}
    \label{thm:main}
    There exists data structures with the following operations.
    Each data structure initializes in $\Otil(n^\omega s \log (\kappa/\epsilon))$ time on given accuracy parameters $\epsilon>0,\kappa > n$, matrix formula $f(\mm_1,...,\mm_s)$, and respective input matrices. Here $n$ is the sum of the number of rows and columns of all $\mm_1,...,\mm_s$, and $\|\mm_i\|_F\le \kappa$ for all $i$, and we assume the result of each inversion within $f$ also has Frobenius-norm bounded by $\kappa$.

    The data structures have the following update/query operations (each bullet is a different data structure)
    \begin{itemize}
        \item Support entry updates and entry queries in $\Otil(n^{\eeEntryExp} s\log(\kappa/\epsilon))$ time. 
        \item Support entry updates in $\Otil(n^{\neEntryUpdateExp}s \log (\kappa/\epsilon))$ time and entry queries in $O(n^{\neEntryQueryExp} s \log(\kappa/\epsilon))$ time. 
        \item Support column updates and row queries in $\Otil(n^{\neEntryUpdateExp} s\log (\kappa/\epsilon))$ time.
        \item Support rank-1 updates and returning all entries of $f(\mm_1,...,\mm_s)$ in $\Otil(n^2 s\log (\kappa/\epsilon))$ time.
        \item Support column updates and row queries in the offline model (the entire sequence of column indices and row queries is given at the start) in $\Otil(n^{\omega-1} s \log(\kappa/\epsilon))$ update and query time. 
    \end{itemize}
    The outputs are all $\epsilon$-approximate, i.e.~each entry is off by at most an additive $\epsilon$.
\end{restatable}

A similar result was previously proven in \cite{Brand21} using data structures from \cite{Sankowski04,BrandNS19}, assuming $O(1)$ time per arithmetic operation and infinite precision.
We extend this to the word-RAM model by analyzing the stability of this data structure under the fixed-point arithmetic.

In addition, we show that we can also maintain other properties of $f(\mm_1,...,\mm_s)$ while receiving updates to the input matrices.
We can maintain the determinant and the rank of $f(\mm_1,...,\mm_s)$.

\begin{restatable}{theorem}{thmDynDet}
\label{thm:dynamic_det}
Let $\mm_1 \in\R^{n_1\times d_1},...,\mm_s\in\R^{n_s\times d_s}$ and $n= \sum_{i=1}^s n_i+d_i$.
    There exists a dynamic determinant data structures that initialize in $\Otil(n^\omega s \log (\kappa/\epsilon))$ time on given accuracy parameters $\epsilon>0,\kappa > 2n$, matrix formula $f(\mm_1,...,\mm_s)$, and respective input matrices.

    The data structures support the maintenance of $\det(f(\mm_1,...,\mm_s))$ up to a multiplicative factor of $1\pm \epsilon$. 
    They have the following update operations (each bullet is a different data structure)
    \begin{itemize}
        \item Support entry updates to any $\mm_i$ in $\Otil(n^{\eeEntryExp} s\log(\kappa/\epsilon))$ time. 
        \item Support column updates to any $\mm_i$ in $\Otil(n^{\neEntryUpdateExp} s \log(\kappa/\epsilon))$ time.
        \item Support rank-1 updates to any $\mm_i$ in $\Otil(n^2 s\log (\kappa/\epsilon))$ time.
    \end{itemize}
    
    We assume that throughout all updates,
    $\|f(\mm_1,...,\mm_s)\|_F \le \kappa$, $\|(f(\mm_1,...,\mm_s))^{-1}\|_F \le \kappa$,
    and $\|\mm_i\|_F\le \kappa$ for all $i$, 
    and the result of each inversion within $f$ also has the Frobenius norm bounded by $\kappa$.
\end{restatable}

\begin{restatable}{theorem}{thmDynRank}\label{thm:dynamicRank}
    There exists a dynamic rank data structure that initializes in $O(n^\omega)$ arithmetic operations on given matrix formula $f(\mm_1,...,\mm_s)$, and respective input matrices. Here, $n$ is the sum of the number of rows and columns of all $\mm_1,...,\mm_s$. The data structure maintains $\rank(f(\mm_1,...,\mm_s))$ subject to entry updates to any $\mm_i$ in $O(n^{\eeEntryExp})$ arithmetic operations per update.
\end{restatable}

This implies, for example, maintaining the size of the maximum matching in a dynamic graph undergoing edge insertions and deletions, turning vertices on/off, and also merging of vertices (\Cref{thm:matching}). Each such update to the graph takes $O(n^{\eeEntryExp})$ time.
This was previously achieved for edge insertions/deletions only \cite{10.5555/1283383.1283397,BrandNS19}.

\subsection{Preliminaries}

\paragraph{Notation} 
We denote matrices with bold uppercase letters and vectors with bold lowercase letters. 
We denote the Frobenius norm and the operator norm by $\norm{\cdot}_{\fro}$ and $\norm{\cdot}_2$, respectively. 
We define the condition number of an invertible matrix $\mM$ as $\kappa(\mM):=\norm{\mM}_2 \cdot\norm{\mM^{-1}}_2$. When the corresponding matrix is clear from the context, we drop the argument and simply write $\kappa$.
We denote entry $(i,j)$ of $\mM$ by $\mM_{i,j}$, row $i$ of $\mM$ by $\mM_{i:}$ and column $j$ of $\mM$ by $\mM_{:j}$. 
For sets $I$ and $J$, we write $(\mA)_{I,}$ to denote the rows with indices in $I$, $(\mA)_{,J}$ to denote the column with indices in $J$, and $(\mA)_{I,J}$ to denote the submatrix with rows with indices in $I$ and columns with indices in $J$.

We denote the $n\times n$ identity matrix by $\mI^{(n)}$, and use $\mathbf{0}^{(i,j)}$ to denote the $i\times j$ all-zeros matrix. 
We denote the transposition of matrix $\mM$ by $\mM^\top$. 
We use $\tilde{O}$ notation to omit polylogarithmic factors in $n$ and polyloglog factors in $\kappa/\epsilon$ from the complexity, i.e., for function $f$, $\tilde{O}(f) := O(f\cdot (\log n \cdot \log\log \frac{\kappa}{\epsilon})^c)$, where $c$ is a constant. 
We denote the set $\{1,\dots,n\}$ by $[n]$. 
We denote the number of operations for multiplying an $n^a\times n^b$ matrix with an $n^b\times n^c$ matrix by $O(n^{\omega(a,b,c)})$ and use $O(n^\omega)$ as shorthand for $O(n^{\omega(1,1,1)})$. 
Finally, for $\mA \in \R^{m\times n}$ and $i\in [\min(m,n)]$, let $\sigma_i$ denote the $i$'th singular value of $\mA$.

\paragraph{Sherman-Morrison-Woodbury Identity \cite{woodbury1950inverting}.}
Consider an invertible matrix $\mM \in \R^{n\times n}$, and matrices $\mU\in \R^{n\times r}, \mD \in \R^{r\times r}, \mV \in \R^{r\times n}$. If $\mD$ and $(\mM + \mU\mD\mV)^{-1}$ are invertible, then:
\begin{align}
(\mM + \mU\mD\mV)^{-1} = \mM^{-1} - \mM^{-1}\mU(\mD^{-1} + \mV\mM^{-1}\mU)^{-1}\mV\mM^{-1} \label{eq:woodbury}
\end{align}

\paragraph{Schur complement.} Consider the block matrix $\mM$ given by:
\[\mm = \begin{bmatrix}
    \ma & \mb \\ \mc & \md
\end{bmatrix},\]
where $\mA$ and $\mD$ are square matrices. Then if $\md$ is invertible, $\mm/\md:= \ma - \mb \md^{-1} \mc$ is called the \emph{Schur complement} of block $\md$ of matrix $\mm$. Similarly, if $\ma$ is invertible, $\mm/\ma:= \md - \mc \ma^{-1} \mb$ is the Schur complement of block $\ma$ of $\mm$. Schur complement gives a nice inversion formula for block matrices. In particular, we have the following. 

\begin{fact}
\label{fact:block-matrix-inverse}
If $\ma$ and $\mm/\ma$ are invertible, then $\mm$ is invertible and
\[
\mm^{-1} = \begin{bmatrix}
    \ma^{-1} + \ma^{-1} \mb (\mm/\ma)^{-1}\mc \ma^{-1} & - \ma^{-1} \mb (\mm/\ma)^{-1} \\
    (\mm/\ma)^{-1}\mc \ma^{-1} & (\mm/\ma)^{-1}
\end{bmatrix}.
\]
\end{fact}
\noindent
This can be easily verified by multiplication with $\mm$.\\

\noindent The Frobenius norm over $\R$ satisfies non-negativity, homogeneity, and the triangle inequality. Specifically, we have that $\norm{\mA}_\fro \geq 0$, $\norm{\beta\mA}_{\fro}=|\beta|\norm{\mA}_{\fro}$ and $\norm{\ma+\mb}_{\fro}\le \norm{\ma}_{\fro}+\norm{\mb}_{\fro}$. Finally, when the product $\mA\mB$ is defined, the Frobenius norm obeys the following submultiplicative property: $\norm{\ma\mb}_{\fro}\le \norm{\ma}_{\fro}\norm{\mb}_{\fro}$.

\paragraph{Our Computational Model.}
Our algorithms and analysis are under the fixed-point arithmetic. We present all of our analysis under fixed-point arithmetic except for the result of \cite{DemmelDH07} for QR decomposition which is under floating-point arithmetic but we only use that result in a black-box way. In fixed-point arithmetic, each number is represented with a fixed number of bits before and after the decimal point, e.g., under fixed-point arithmetic with $6$ bits, we can only present integer numbers less than $64$. Addition/subtraction and multiplication of two numbers with $n$ bits, can be done in $\Otil(n)$ times in this model by using fast Fourier transform (FFT) --- see \cite[Chapter~30]{cormen2022introduction}. Division to an additive error of $\epsilon$ can also be performed in $\Otil(n+\log(1/\epsilon))$ again with the help of FFT. In general, when we mention running time, we mean the number of bit operations. Otherwise, we specify the complexity is about the number of arithmetic operations.

\paragraph{Matrix Formula}

Intuitively, a matrix formula is any formula involving matrices and the basic matrix operations of adding, subtracting, multiplying, or inverting matrices. E.g., $f(\mA,\mB,\mC,\mD) = (\mA \mB + \mC)^{-1} \mD$ is a matrix formula.

Formally, a matrix formula $f(\mm_1,...,\mm_s)$ is a directed tree, where each input $\mm_i$ is a leaf, and each matrix operation (addition, subtraction, multiplication, inversion) is an internal node. Nodes that represent addition, subtraction, or multiplication have two children, i.e.~the two terms that are being added, subtracted, or multiplied. Inversion has only one child, the term being inverted.
For example, for node $v$ labeled ``$+$'', the subtree rooted at the left child and the subtree rooted at the right child represent formulas $g_{\textnormal{left}}(\mm_1,...,\mm_k), g_{\textnormal{right}}(\mm_{k+1},...,\mm_s)$, and the tree rooted at $v$ represents $f(\mm_1,...,\mm_s) = g_{\textnormal{left}}(\mm_1,...,\mm_k) + g_{\textnormal{right}}(\mm_{k+1},...,\mm_s)$.

Note that since a formula is a tree, and not a DAG, and because there is no point in inverting something twice in succession, a formula (i.e., tree) with $s$ input matrices (i.e., leafs) has at most $O(s)$ operations (i.e., internal nodes).
Further, note that by formulas being trees, a leaf (input) can be used only once. For example, $(\mA+\mB)\mA$ is a formula with 3 inputs.

\section{Technical Overview}

Here we outline how we obtain our three main results \Cref{thm:main,thm:dynamic_det,thm:dynamicRank}.
\cite{Brand21} proved a variant of \Cref{thm:main} that assumed the ``real-RAM model'', i.e., exact arithmetic in only $O(1)$ time per operation.
Modern computers do not provide this guarantee unless one uses up to $\Omega(n)$ bit, substantially slowing down each arithmetic operation and the overall algorithm.
We show that the techniques by \cite{Brand21} also works with bounded accuracy and much smaller bit-complexity. 
In particular, only $\tilde O(s \log \kappa)$ bit are enough, as stated in \Cref{thm:main}.

Before we can outline how to obtain these results, we need to give a brief recap of \cite{Brand21}. This is done in \Cref{sec:overview:previous}.
We outline in \Cref{sec:overview:frobenius} how to prove that $\tilde O(s \log \kappa)$-bit accuracy suffice, resulting in \Cref{thm:main}.
At last, \Cref{sec:overview:determinant} outlines how to extend \Cref{thm:main} to maintain determinant and rank of a matrix formula, i.e., prove \Cref{thm:dynamic_det,thm:dynamicRank}.

\subsection{Setting the Stage: How to Maintain Dynamic Algebraic Formulas}
    \label{sec:overview:previous}
    \emph{Dynamic Matrix Formula} is the following data structures task:
    We are given a formula $f(\mm_s,...,\mm_s)$ and respective input matrices $\mm_1,...,\mm_s$. 
    The entries of these matrices change over time, and the data structure must maintain $f(\mm_1,...,\mm_s)$ (see \Cref{thm:main}).
    \emph{Dynamic Matrix Inverse} is the special case $f(\mm) = \mm^{-1}$, i.e., given a matrix that changes over time, we must maintain information about its inverse.
    The latter problem has been studied in, e.g., \cite{Sankowski04,BrandNS19,lin2023online2} and there exist several data structures for this task (see \Cref{thm:inverseds}).

\begin{figure}[t]
        \centering
        \includegraphics[clip, trim=6.5cm 8.5cm 13.5cm 4.5cm, scale=0.5]{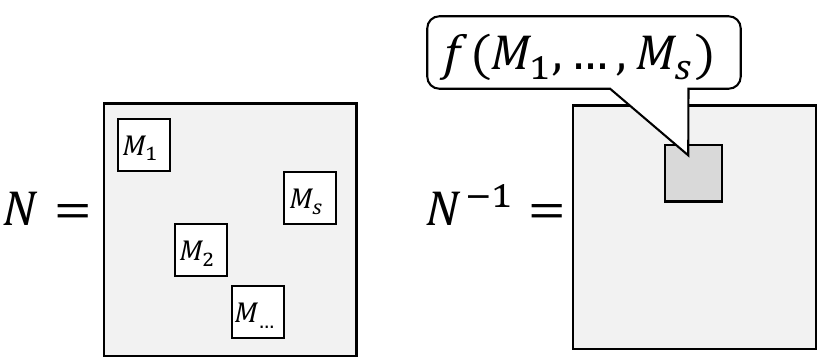}
        \caption{Maintaining $\mn^{-1}$ allows us to maintain $f(\mm_1,...,\mm_s)$. \label{fig:reduction}}
        \label{fig:enter-label}
    \end{figure}

    \cite{Brand21} showed that dynamic matrix formula for any formula can be reduced to the special case of matrix inversion, i.e., dynamic matrix inverse. 
    In particular, this means all the previous data structures for dynamic matrix inverse \cite{Sankowski04,BrandNS19} or simple application of the Sherman-Morrison-Woodbury identity \eqref{eq:woodbury}, can also be used to maintain any general formula $f(\mm_1,...,\mm_s)$.
    \cite{Brand21} shows that for any formula $f(\mm_1,...,\mm_s)$, there is large block matrix $\mn$, where $\mm_1,...,\mm_s$ occur as subblocks.
    Further, $\mn^{-1}$ contains a block\footnote{The proof is constructive. Given $f,\mm_1,...,\mm_s$, we know $\mn$ and we know which submatrix of $\mn^{-1}$ contains $f(\mm_1,...,\mm_s)$, i.e., we do not have to search for the subblock.} that is precisely $f(\mm_1,...,\mm_s)$. See \Cref{fig:reduction}.
    When $\mm_1,...,\mm_s$ change over time, matrix $\mn$ changes over time, and running a dynamic matrix inverse data structures on this $\mn$ allows us to maintain $\mn^{-1}$ and its subblock containing $f(\mm_1,...\mm_s)$.

    The issue of the reduction is that we do not know if $\mn$ is well-conditioned. 
    Under which conditions to $f$ and $\mm_1,...,\mm_s$ can we guarantee that matrix $\mn$ is well-conditioned?
    Once we can guarantee that $\mn$ is well-conditioned, we can give good error guarantees for the dynamic matrix inverse data structures that maintain $\mn^{-1}$ via the result of \cite{MehrdadPV23} regarding the numerical stability of SMW identity (see \cref{lem:woodbury_error}).
    
    We will bound both $\|\mn\|_{\fro}$ and $\|\mn^{-1}\|_{\fro}$, which gives a bound on the condition number.
    In particular, we show that under reasonable assumptions on formula $f$ and input $\mm_1,...,\mm_s$, both Frobenius-norms are bounded by $\kappa^{O(s)}$ (\Cref{lemma:n-cond-bound}).
    With the dynamic matrix inverse data structures' complexities scaling in the log of these Frobenius norms (see \Cref{thm:inverseds}), this leads to the $\tilde O(s \log \kappa)$ factors in \Cref{thm:main}.

\subsection{Bounding the Frobenius Norms}
\label{sec:overview:frobenius}
As outlined in the previous subsection, when given a matrix formula $f(\mm_1,...,\mm_s)$, \cite{Brand21} constructs a matrix $\mn$ with the following properties.
Matrix $\mn$ contains $\mm_1,...,\mm_s$ as subblock, and the inverse $\mn^{-1}$ contains a subblock that is precisely $f(\mm_1,...,\mm_s)$, see \Cref{fig:reduction}.
Our task is to bound the Frobenius-norm of both $\mn$ and $\mn^{-1}$, which then implies a bound on the condition number of $\mn$. Further, data structures for maintaining $\mn^{-1}$ have a runtime that scales in those norms (\Cref{thm:inverseds}).

Let us briefly recap how matrix $\mn$ is constructed, so we can then analyze the Frobenius-norms of $\mn$ and $\mn^{-1}$.

\paragraph{Construction of $\mn$}
    
    The given formula $f(\mm_1,...,\mm_s)$ can be represented as a tree (where the operations like matrix product, sum, or inversion are nodes, the input matrices are leaves, and the output is the root).
    The construction of $\mn$ follows by induction over the size of the tree:
    E.g., given some $f(\mm_1,...,\mm_s) = g_1(\mm_1,...,\mm_k) \cdot g_2(\mm_{k+1},...,\mm_s)$,
    by induction hypothesis there are matrices $\mn_1, \mn_2$ where $\mn_i$ contains the input matrices of $f_i$ as blocks, and a block of $\mn_i^{-1}$ contains the evaluation of $f_i$.
    These two matrices are then combined into one larger matrix $\mn$ (i.e., $\mn$ contains $\mn_1,\mn_2$ as subblocks and thus $\mn$ contains $\mm_1,...,\mm_s$ as subblocks) with the property that $\mn^{-1}$ contains a subblock that is precisely $g_1(\mm_1,...,\mm_k) \cdot g_2(\mm_{k+1},...\mm_s)$.
    A full description of the reduction is given in \Cref{lemma:matrix-formula} for completeness sake.
    To follow the outline, it is only important to know that we perform induction by splitting the formula $f$ at its root into $g_1$ and $g_2$ to obtain two smaller matrices $\mn_1,\mn_2$. 
    If the root is an inversion, i.e., $f(\mm_1,...,\mm_s) = (g(\mm_1,...,\mm_s))^{-1}$, then we only have one matrix $\mn'$ where $\mn'^{-1}$ contains a subblock that is $g(\mm_1,...,\mm_s)$.

\paragraph{Bounding the Frobenius Norm (\Cref{sec:frobeniusnorm})}

For each possible operation $\circ \in \{+,-,\cdot\}$ at the root: $f(\mm_1,...,\mm_s) = g_1(\mm_1,...,\mm_k) \circ g_2(\mm_{k+1},...,\mm_s)$ there is a different construction for how to combine $\mn_1,\mn_2$ into a single $\mn$, such that $\mn^{-1}$ contains $f(\mm_1,...,\mm_s)$ as a submatrix.
We bound the Frobenius-norm of $\mn$ and $\mn^{-1}$, with respect to the Frobenius-norms of $\mn_1,\mn_2$. This then implies a bound by induction.
Since the construction of $\mn$ differs depending on the operation $\circ \in \{+,-,\cdot\}$, we need slightly different proof for each operation.
The proofs will all follow via simple applications of triangle-inequalities. 
Since $\mn$ is constructed in such a way that $\mn_1,\mn_2$ are submatrices of $\mn$, simple arguments via triangle-inequality suffice.
However, for the special case of $f(\mm_1,...,\mm_s) = (g(\mm_1,...,\mm_s))^{-1}$, a more careful analysis is required, which we outline below.

For $\mn'$ being the matrix constructed for formula $g(\mm_1,...,\mm_s)$, we have sets $I',J' \subset \N$ where $(\mn'^{-1})_{I',J'} = g(\mm_1,...,\mm_s)$ (i.e., this is the subblock that contains the evaluation of $g(\mm_1,...,\mm_s)$). 
The reduction by \cite{Brand21} then constructs $\mn$ as follows, and we state its inverse:
\begin{align*}
    \mN=\begin{bmatrix}
        \mN'&-\mI^{(n_{N'})}_{,J'}\\
        \mI^{(n_{N'})}_{I',}&\mzero^{(n_w,n_w)}
    \end{bmatrix}
    \quad\quad\quad
    \mN^{-1}=\begin{bmatrix}
        \mN'^{-1}-(\mN'^{-1})_{,J'}(\mN'^{-1})_{I',J'}^{-1}(\mN'^{-1})_{I',}& (\mN'^{-1})_{,J'}(\mN'^{-1})_{I',J'}^{-1}\\
        -(\mN'^{-1})_{I',J'}^{-1}(\mN'^{-1})_{I',}&((\mN'^{-1})_{I',J'})^{-1}
    \end{bmatrix}
\end{align*}
Note that the bottom right block of $\mn^{-1}$ is precisely $g(\mm_1,...,\mm_s)^{-1}$ since $(\mn'^{-1})_{I',J'} = g(\mm_1,...,\mm_s)$.
To bound the Frobenius-norm of $\mn$ and $\mn^{-1}$, we apply the triangle inequality to
\begin{align*}
    \norm{\mN}_{\fro}\le  \norm{\begin{bmatrix}
        \mN'&0\\
        0&0
        \end{bmatrix}}_{\fro}+\norm{\begin{bmatrix}
        0&-\mI_{,J'}^{(n_{N'})}\\
        \mI_{I',}^{(n_{N'})}&0
        \end{bmatrix}}_{\fro}
        \le
        \|\mn'\|_F + \sqrt{|J'|+|I'|}
\end{align*}
We will have an upper bound on $\|\mN'\|_{\fro}$ by the inductive hypothesis, so this yields an upper bound on $\|\mN\|_{\fro}$ as well.

Using the triangle inequality similarly, we also can also bound $\|\mN^{-1}\|_{\fro}$ by
\begin{align}
    \|\mN^{-1}\|_{\fro} 
    & 
    \le \norm{\mN'^{-1}-(\mN'^{-1})_{,J'}(\mN'^{-1})_{I',J'}^{-1}(\mN'^{-1})_{I',}}_{\fro}
    \notag\\ & +
    \norm{(\mN'^{-1})_{,J'}(\mN'^{-1})_{I',J'}^{-1}}_{\fro}+      \norm{(\mN'^{-1})_{I',J'}^{-1}(\mN'^{-1})_{I',}}_{\fro}+\norm{(\mN'^{-1})_{I',J'}^{-1}}_{\fro} \label{eq:overview:N}
\end{align}

We can split the sum by the triangle inequality, and products using $\|\mA\mB\|_{\fro}\le \|\mA\|_{\fro}\|\mB\|_{\fro}$.

If we naively upper bound $\|(\mN'^{-1})_{I',}\|_{\fro}$ and $\|(\mN'^{-1})_{,J'}\|_{\fro}$ using $\|(\mN'^{-1})\|_{\fro}$, we will get

\begin{align*}
\|\mN'^{-1}-(\mN'^{-1})_{,J'}(\mN'^{-1})_{I',J'}^{-1}(\mN'^{-1})_{I',}\|_{\fro}&\le \|\mN'^{-1}\|_{\fro}+\|(\mN'^{-1})_{,J'}\|_{\fro}\|(\mN'^{-1})_{I',J'}^{-1}\|_{\fro}\|(\mN'^{-1})_{I',}\|_{\fro}\\
&\le \|\mN'^{-1}\|_{\fro}+\|(\mN'^{-1})_{I',J'}^{-1}\|_{\fro}\|(\mN'^{-1})\|_{\fro}^2
\end{align*}
and similarly for the other three terms of \eqref{eq:overview:N}. This will yield an upper bound for $\|\mN^{-1}\|_{\fro}$, but it involves $\|(\mN'^{-1})\|_{\fro}^2$. In particular, the upper bound gets squared for every nested inverse gate, which will yield a bound that is in the order of $\kappa^{2^s}$ (with $O(s)$ being a bound on the number of gates).

To improve this, we bound $\|(\mN'^{-1})_{I',}\|_{\fro}$ and $\|(\mN'^{-1})_{,J'}\|_{\fro}$ inductively as well. This removes the dependence on $\|(\mN'^{-1})\|_{\fro}^2$, so the upper bound no longer gets squared in every iteration, and becomes $\kappa^{O(s)}$ instead.

\subsection{Dynamic Rank and Determinant of Matrix Formulas}
\label{sec:overview:determinant}

So far, we outlined how to maintain $f(\mm_1,...\mm_s)$ within finite precision. This is based on a reduction by \cite{Brand21} from dynamic matrix formula to dynamic matrix inverse.
We now explain how to extend the reduction, allowing us to also maintain $\det(f(\mm_1,...,\mm_s))$ and $\rank(f(\mm_1,...,\mm_s))$.

\paragraph{Maintaining the Determinant (\Cref{sec:determinant})}
First, note that given a block matrix, we can represent its determinant as follows
\[
\text{For }
\mm = \begin{bmatrix}
    \ma & \mb \\
    \mc^\top & \md
\end{bmatrix}
\text{ we have }
\det(\mm) = \det(\ma) \cdot\det(\md - \mc^\top \ma^{-1} \mb).
\]
This allows for the following observation: Given $n\times n$ matrix $\mn$ and sets $I,J \subset \N$ with $(\mn^{-1})_{I,J} = f(\mm_1,...,\mm_s)$, 
we have $(\mn^{-1})_{I,J}=\mI^{(n)}_{I,[n]} \mn^{-1} \mI^{(n)}_{[n],J}$ and thus
\[
\hat\mN = \begin{bmatrix}
    \mn & \mI^{(n)}_{[n],J} \\
    \mI^{(n)}_{I,[n]} & \mathbf{0}
\end{bmatrix}
\text{ with }
\det(f(\mm_1,...,\mm_s))
=
\det(\mI^{(n)}_{I,[n]} \mn^{-1} \mI^{(n)}_{[n],J})
=
\det(\hat\mn)/\det(\mn).
\]
Thus, to maintain $\det(f(\mm_1,...,\mm_s))$, we just need to maintain $\det(\hat\mn)$ and $\det(\mn)$.
Maintaining these determinants can be done via the determinant lemma, which states:
\begin{align}
\det(\mn+\mathbf{u}\mathbf{v}^\top) = \det(\mn)(1 + \mathbf{v}^\top \mn^{-1} \mathbf{u}). \label{eq:determinantlemma}
\end{align}
Here adding $\mathbf{u}\mathbf{v}^\top$ is a rank-1 update, and can capture updates such as changing an entry of $\mn$ (when $\mathbf{u}, \mathbf{v}$ have only 1 nonzero entry each) or changing a column of $\mn$ (when $v$ has only 1 nonzero entry).
In particular, the task of maintaining $\det(\mn)$ reduces to the task of repeatedly computing $\mathbf{v}^\top \mn^{-1} \mathbf{u}$.
This is a dynamic matrix formula (since $\mathbf{u},\mathbf{v},\mn$ change over time). For example, maintaining $\det(\mn)$ while $\mn$ receives entry updates, requires us to maintain $f(\mathbf{u},\mathbf{v},\mn) = \mathbf{v}^\top \mn^{-1} \mathbf{u}$ while $\mathbf{u},\mathbf{v},\mn$ receive entry updates.
Thus data structures for maintaining $\det(f(\mm_1,...,\mm_s))$ (\Cref{thm:dynamic_det}) are implied by data structures for dynamic matrix formula (\Cref{thm:inverseds}), together with some additional error analysis performed in \cref{sec:determinant}. A key observation in the error analysis is that the determinant is the product of the eigenvalues, and therefore, if we guarantee (with a sufficient number of bits) that the eigenvalues are preserved up to a multiplicative error factor of $1\pm \frac{\epsilon}{10n}$, then we have determinant computation up to a multiplicative error factor of $1\pm \epsilon$. We formalize this idea by bounding the determinant of a matrix perturbed by a small amount --- see \cref{lemma:sum-det}.

\paragraph{Maintaining the Rank (\Cref{sec:rank})}

Let us assume that $\mm_1,...,\mm_s$ are integer matrices, so $\mn$ is an integer matrix as well.
Note that whp.~$\rank(\mn)$ is the same over $\Z$ and $\Z_p$ for prime $p \sim n^c$ and large enough constant $c$.
So for the rank, we do not need to worry about rounding errors and can just focus on finite fields.

Sankowski \cite{10.5555/1283383.1283397} proved the following statement about matrix ranks.
For any $n\times n$ matrix $\mm$, and random $n\times n$ matrices $\mX$ and $\mY$ (each entry is chosen uniformly at random from $\Z_p$), and $\mI_k$ being a partial identity (the first $k$ diagonal entries are $1$, the remaining diagonal entries are $0$), let
\[\overline{\mathbf{M}} = \begin{bmatrix}
        \mathbf{M} & \mathbf{X} & \mathbf{0} \\
        \mathbf{Y} & \mathbf{0} & \mathbf{I}_n \\
        \mathbf{0} & \mathbf{I}_n & \mathbf{I}_k
    \end{bmatrix}\] 
Then with high probability, ~$\det(\overline{\mathbf{M}}) \neq 0 \iff \rank(\mathbf{M})\geq n-k$.
In \cite{10.5555/1283383.1283397}, this was used to maintain the rank of $\mm$. We now generalize this to maintaining the rank of $f(\mm_1,...,\mm_s)$.\\

\noindent Given a formula $f(\mm_1,...,\mm_s)$, let $g(\mm_1,...,\mm_s,\mP,\mQ,\mR_k) =\mP f(\mm_1,...,\mm_s) \mQ+\mR_k$
where
\begin{align*}
    \mP = \mQ = \begin{bmatrix}
        \mI & \mathbf{0} & \mathbf{0} \\
        \mathbf{0} & \mathbf{0} & \mathbf{0} \\
        \mathbf{0} & \mathbf{0} & \mathbf{0} \\
    \end{bmatrix},
    \quad\quad
    \quad\quad
    \mR_k = \begin{bmatrix}
        \mathbf{0} & \mX & \mathbf{0} \\
        \mY & \mathbf{0} & \mI \\
        \mathbf{0} & \mI & \mI_k \\
    \end{bmatrix}
\end{align*}
we have
\[
g(\mm_1,...,\mm_s,\mP,\mQ,\mR_k)
 = \begin{bmatrix}
        f(\mm_1,...,\mm_s) & \mathbf{X} & \mathbf{0} \\
        \mathbf{Y} & \mathbf{0} & \mathbf{I}_n \\
        \mathbf{0} & \mathbf{I}_n & \mathbf{I}_k
    \end{bmatrix}\] 
Thus, $\det(g(\mm_1,...,\mm_s,\mP,\mQ,\mR_k)) \neq 0 \iff \rank(f(\mm_1,...,\mm_s))\geq n-k$.
So we can track the rank of $f(\mm_1,...,\mm_s)$ by finding and maintaining the smallest $k$ where $\det(g(\mm_1,...,\mm_s,\mP,\mQ,\mR_k)) \neq 0$.
Note that with each changed entry to any $\mm_i$, the rank can change by at most $1$. 
So we can simply try increasing/decreasing $k$ by performing a single entry update to $\mR_k$ (and potentially reverting the update) to check if the $\det(g(\mm_1,...,\mm_s,\mP,\mQ,\mR_k))$ becomes $0$. 
Note that the determinant lemma \eqref{eq:determinantlemma} breaks once the matrix is no longer invertible. Thus we must increase $k$ whenever $\det(g(\mm_1,...,\mm_s,\mP,\mQ,\mR_k))=0$.
If an update causes the determinant to become $0$, we must revert that update by reverting any internal changes of the data structure, then increase $k$, and then perform the reverted update again. This way, we always guarantee $\det(g(\mm_1,...,\mm_s,\mP,\mQ,\mR_k))\neq0$ and that \eqref{eq:determinantlemma} never breaks.
By always maintaining the largest $k$ where $\det(g(\mm_1,...,\mm_s,\mP,\mQ,\mR_k)) \neq 0$, we know $n-k$ is the rank of $f(\mm_1,...,\mm_s)$.
\section{Dynamic Matrix Formula}

In this section we prove the first main result. Generic data structure that can maintain the evaluation of any matrix formula $f(\mm_1,...,\mm_s)$ while supporting updates to the input matrices.

\thmFormulaMaintenance*

This result is obtained by reducing the task to the special case $g(\mn) = \mn^{-1}$, where the structure of matrix $\mn$ depends on the formula $f$ and its inputs $\mm_1,...,\mm_s$.
This reduction is given by \Cref{lemma:matrix-formula}.

We then run data structures (\Cref{thm:inverseds}) that can maintain $\mn^{-1}$ while supporting updates to $\mn$.
The accuracy of these data structures depends on $\|\mn\|_F$ and $\|\mn^{-1}\|_F$, so we must bound these Frobenius-norms.
These bounds are given in \Cref{sec:frobeniusnorm}.
We then combine the bounds with the previous reduction to obtain \Cref{thm:main} in \Cref{sec:main}.

\subsection{Norm Bounds on Construction}
\label{sec:frobeniusnorm}
In this section, we bound the Frobenius norm of the matrix produced in \Cref{lemma:matrix-formula}, as well as its inverse.

\begin{lemma}
\label{lemma:n-cond-bound}
    Let $f(\mA_1,\ldots,\mA_p)$ be a matrix formula over $\R$ using $s$ gates. Suppose matrix $\mN$, and index sets $I$, and $J$ are constructed as in \Cref{lemma:matrix-formula} so that $(\mN^{-1})_{I,J}=f(\mA_1,\ldots,\mA_p)$.
    Let $\kappa\ge \max_{i}n_i+m_i\ge 2$, where $n_i\times m_i$ are the dimensions of $\mA_i$.
    
    Then, if $\|\mA_1\|_{\fro},\ldots,\|\mA_p\|_{\fro}\le \kappa$, and the Frobenius norms of outputs of intermediate inverse gates are also bounded by $\kappa$, we have
    \begin{align*}
        \|\mN\|_{\fro} \le \kappa^s, ~~ \text{and} ~~
        \|\mN^{-1}\|_{\fro} \le (10\kappa)^{2s+1}.
    \end{align*}
\end{lemma}

\begin{proof}
    We bound $\|\mN\|_{\fro}$ and $\|\mN^{-1}\|_{\fro}$ by induction on the number of gates $s$. 
    We write $\mN$ and $\mN^{-1}$ as block matrices in the same way that they were constructed --- see \cref{lemma:matrix-formula}.
    Note that the given formula $f$ can be represented as a tree, where the input matrices are leaves and each operation is an internal node. Each node that represents an operation has 2 children (or 1 child in case of inversion). We call the nodes also gates, e.g., inversion gate or addition gate, depending on what operation they represent. 
    \Cref{lemma:matrix-formula} construct the matrix $\mN$ by induction over the number of gates, i.e., for each gate $w$ some matrix $\mn_w$ is constructed. This $\mn_w$ is constructed as a block matrix where some blocks are $\mn_u,\mn_v$ where $u,v$ are the child gates of $w$. We also say that ``$\mn_w$ is returned by gate $w$''.

    We can bound the Frobenius norm of a block matrix by the sum of the Frobenius norms of its blocks, using the triangle inequality to write the matrix as a sum of matrices where all but one (or a few) block is zero. For $\mN^{-1}$, we then use the inequalities $\|\mA+\mB\|_{\fro}\le \|\mA\|_{\fro}+\|\mB\|_{\fro}$ and $\|\mA\mB\|_{\fro}\le \|\mA\|_{\fro}\|\mB\|_{\fro}$ to bound the Frobenius norms of the expressions in each block.

    If we do this naively, we get a $\kappa^{2^s}$ term in the bound since the $\mN'^{-1}-(\mN'^{-1})_{,J'}(\mN'^{-1})_{I',J'}^{-1}(\mN'^{-1})_{I',}=\mN'^{-1}-(\mN'^{-1})\mI^{(n_{N'})}_{,J'}\mI^{(n_{N'})}_{I',}(\mN'^{-1})^{-1}\mI^{(n_{N'})}_{,J'}\mI^{(n_{N'})}_{I',}(\mN'^{-1})$ block in the inversion gate causes the bound on the Frobenius norm to be squared each time. To avoid this, we first separately bound the $\|(\mN^{-1})_{I,}\|_{\fro}$ and $\|(\mN^{-1})_{,J}\|_{\fro}$ terms by induction, and then use them to bound $\|\mN^{-1}\|_{\fro}$. More precisely, we will show by induction on $s\ge 1$ that:
    \begin{align*}
        \|\mN\|_{\fro}&\le \kappa^s&\|(\mN^{-1})_{I,J}\|_{\fro}&\le 2s\kappa &\|(\mN^{-1})_{I,}\|_{\fro}&\le (5\kappa)^s&
        \|(\mN^{-1})_{,J}\|_{\fro}&\le (5\kappa)^s&\|\mN^{-1}\|_{\fro}&\le (10\kappa)^{2s+1}
    \end{align*}

We now prove bounds on the output of our gates by assuming the induction hypothesis that their inputs have bounded norms. We start with the base case.
\paragraph{Input gate.}
    The base case is when $s=1$, in which case the formula consists of a single input gate. We have
\begin{align*}
    \mN^{-1}=\mN&=\begin{bmatrix}
        \mI^{(n_v)}&\mM_v\\
        \mzero^{(m_v,n_v)}&-\mI^{(m_v)}
    \end{bmatrix}
\end{align*}
where $\mM_v$ is the input matrix and $n_v \times m_v$ are its dimensions. 
Selecting rows with indices in $I=\{1,...,m_v\}$, and columns with indices in $J=\{n_v+1,...,n_v+m_v\}$, we get
\begin{align*}(\mN^{-1})_{,J}&=\begin{bmatrix}
        \mM_v\\
        -\mI^{(m_v)}
    \end{bmatrix}, &
    (\mN^{-1})_{I,}&=\begin{bmatrix}
        \mI^{(n_v)}&\mM_v
    \end{bmatrix},&
    (\mN^{-1})_{I,J}&=\begin{bmatrix}
        \mM_v
    \end{bmatrix}
\end{align*}
\noindent
Applying the triangle inequality to
\begin{align*}
    \mN=\begin{bmatrix}
        \mI^{(n_v)}&\mzero^{(n_v,m_v)}\\
        \mzero^{(m_v,n_v)}&-\mI^{(m_v)}
    \end{bmatrix}+\begin{bmatrix}
        \mzero^{(n_v,n_v)}&\mM_v\\
        \mzero^{(m_v,n_v)}&\mzero^{(m_v,m_v)}
    \end{bmatrix}
\end{align*}
and using the fact that $\sqrt{n_v+m_v}\le \kappa$ and $\|\mM_v\|_{\fro}\le \kappa$, we get
\begin{align*}
    \|\mN\|_{\fro}&\le \sqrt{n_v+m_v}+\|\mM_v\|_{\fro}\le 2\kappa
\end{align*}
Similarly, we have that
\begin{align*}
    \|(\mN^{-1})_{I,J}\|&\le \|\mM_v\|_{\fro}\le \kappa^1\\
    \|(\mN^{-1})_{I,}\|&\le \|\mM_v\|_{\fro}+\sqrt{n_v}\le 2\kappa\le (5\kappa)^1\\
    \|(\mN^{-1})_{,J}\|&\le \|\mM_v\|_{\fro}+\sqrt{m_v}\le 2\kappa\le (5\kappa)^1\\
    \|\mN^{-1}\|_{\fro}&\le \sqrt{n_v+m_v}+\|\mM_v\|_{\fro}\le 2\kappa\le (10\kappa)^3
\end{align*}
We now consider the operation gates for the inductive step.
\paragraph{Inversion.}
Suppose the root gate is a inversion gate. Suppose $\mN'$ is a $n_{N'}\times n_{N'}$ matrix and $I',J'\subset \Z$ are sets, such that $(\mN'^{-1})_{I',J'}$ is the $n_w\times n_w$ matrix that the child node $w$ outputs. The child node has $a=s-1\ge 1$ gates.
\begin{align*}
    \mN&=\begin{bmatrix}
        \mN'&-\mI^{(n_{N'})}_{,J'}\\
        \mI^{(n_{N'})}_{I',}&\mzero^{(n_w,n_w)}
    \end{bmatrix}
\end{align*}
By block matrix inversion (\cref{fact:block-matrix-inverse}), we have
\begin{align*}
    \mN^{-1}&=\begin{bmatrix}
        \mN'^{-1}-(\mN'^{-1})_{,J'}(\mN'^{-1})_{I',J'}^{-1}(\mN'^{-1})_{I',}& (\mN'^{-1})_{,J'}(\mN'^{-1})_{I',J'}^{-1}\\
        -(\mN'^{-1})_{I',J'}^{-1}(\mN'^{-1})_{I',}&(\mN'^{-1})_{I',J'}^{-1}
    \end{bmatrix}
\end{align*}
\noindent
Selecting the rows with indices in $I$, and columns with indices in $J$, we get
\begin{align*}(\mN^{-1})_{I,} &= \begin{bmatrix}
        -(\mN'^{-1})_{I',J'}^{-1}(\mN'^{-1})_{I',}&(\mN'^{-1})_{I',J'}^{-1}
    \end{bmatrix}\\
(\mN^{-1})_{,J} &= \begin{bmatrix}
        (\mN'^{-1})_{,J'}(\mN'^{-1})_{I',J'}^{-1}\\
        (\mN'^{-1})_{I',J'}^{-1}
    \end{bmatrix} \\ (\mN^{-1})_{I,J} &= \begin{bmatrix}
(\mN'^{-1})_{I',J'}^{-1}
    \end{bmatrix}
\end{align*}
\noindent
By the triangle inequality and the assumption that $\kappa\ge n_w\ge 1$,
\begin{align*}
    \|\mN\|_{\fro}&\le \|\mN'\|_{\fro}+\sqrt{2n_w}\\
    &\le 2a\kappa+2\kappa\le 2s\kappa
\end{align*}
\noindent
By the assumption that the output of each inversion gate has Frobenius norm at most $\kappa$,
\begin{align*}
    \|(\mN^{-1})_{I,J}\|_{\fro}&=\|(\mN'^{-1})_{I',J'}^{-1}\|_{\fro}\le \kappa\le\kappa^s
\end{align*}

\noindent For the remaining matrices, we can bound their Frobenius norms by the sum of Frobenius norms of their blocks, use the triangle inequality to split sums and $\|\mA\mB\|_{\fro}\le\|\mA\|_{\fro}\|\mB\|_{\fro}$ to split products, and then bound the resulting terms by the inductive hypothesis:
\begin{align*}
    \|(\mN^{-1})_{I,}\|_{\fro}&\le \|(\mN'^{-1})_{I',J'}^{-1}\|_{\fro}(1+\|(\mN'^{-1})_{I',}\|_{\fro})\\
    &\le \kappa(1+(5\kappa)^a)\le \kappa\cdot 2(5\kappa)^a\le (5\kappa)^{a+1}=(5\kappa)^s\\
    \|(\mN^{-1})_{,J}\|_{\fro}&\le \|(\mN'^{-1})_{I',J'}^{-1}\|_{\fro}(1+\|(\mN'^{-1})_{,J'}\|_{\fro})\\
    &\le \kappa(1+(5\kappa)^a)\le \kappa\cdot 2(5\kappa)^a\le (5\kappa)^{a+1}=(5\kappa)^s\\
    \|\mN^{-1}\|_{\fro}&\le \|\mN'^{-1}\|_{\fro}+\|(\mN'^{-1})_{I',J'}^{-1}\|_{\fro}(\|(\mN'^{-1})_{I',}\|_{\fro}+1)(\|(\mN'^{-1})_{,J'}\|_{\fro}+1)\\
    &\le (10\kappa)^{2a+1}+\kappa((5\kappa)^a+1)^2\le (10\kappa)^{2a+1}+\kappa(10\kappa)^{2a}\le (10\kappa)^{2a+3}=(10\kappa)^{2s+1}
\end{align*}
\paragraph{Addition and Subtraction.}
Suppose the root gate $w$ is an addition gate, adding two $n_w \times m_w$ matrices.
Suppose the left child has $a\ge 1$ gates and the right child has $b\ge 1$ gates, where $s=a+b+1$.
Let $\mL$ be the $n_L\times n_L$ matrix and $\mR$ be the a $n_R\times n_R$ matrix such that the outputs of the child gates are $(\mL^{-1})_{I_L,J_L}$ and $(\mR^{-1})_{I_R,J_R}$, which are $n_w\times m_w$ matrices.

\begin{align*}
    \mN&=\begin{bmatrix}
        \mL&\mzero&\mI^{(n_L)}_{,J_L}&\mzero\\
        \mzero&\mR&\mI^{(n_R)}_{,J_R}&\mzero\\
        \mI^{(n_L)}_{I_L,}&\mI^{(n_R)}_{I_R,}&\mzero&\mI^{(n_w)}\\
        \mzero&\mzero&\mI^{(m_w)}&\mzero
    \end{bmatrix}\\
    \mN^{-1}&=\begin{bmatrix}
        \mL^{-1}&\mzero&\mzero&-(\mL^{-1})_{,J_L}\\
        \mzero&\mR^{-1}&\mzero&-(\mR^{-1})_{,J_R}\\
        \mzero&\mzero&\mzero&\mI^{(m_w)}\\
        -(\mL^{-1})_{I_L,}&-(\mR^{-1})_{I_R,}&\mI^{(n_w)}&(\mL^{-1})_{I_L,J_L}+(\mR^{-1})_{I_R,J_R}\\
    \end{bmatrix}
\end{align*}
\noindent
Selecting the rows of $\mN^{-1}$ with indices in $I$ and columns with indices in $J$, we get
\begin{align*}
    (\mN^{-1})_{I,}&=\begin{bmatrix}
        -(\mL^{-1})_{I_L,}&-(\mR^{-1})_{I_R,}&\mI^{(n_w)}&(\mL^{-1})_{I_L,J_L}+(\mR^{-1})_{I_R,J_R}\\
    \end{bmatrix}\\
    (\mN^{-1})_{,J}&=\begin{bmatrix}
        -(\mL^{-1})_{,J_L}\\
        -(\mR^{-1})_{,J_R}\\
        \mI^{(m_w)}\\
        (\mL^{-1})_{I_L,J_L}+(\mR^{-1})_{I_R,J_R}\\
    \end{bmatrix}\\
    (\mN^{-1})_{I,J}&=\begin{bmatrix}
        (\mL^{-1})_{I_L,J_L}+(\mR^{-1})_{I_R,J_R}\\
    \end{bmatrix}
\end{align*}
We now bound the Frobenius norms of these matrices using the bounds on $\|\mL\|_{\fro}$, $\|\mL^{-1}\|_{\fro}$, $\|(\mL^{-1})_{I_L,}\|_{\fro}$, $\|(\mL^{-1})_{,J_L}\|_{\fro}$, $\|(\mL^{-1})_{I_L,J_L}\|_{\fro}$, and similarly for $\mR^{-1}$, that we get from the inductive hypothesis. Using the triangle inequality and the assumption that $\kappa\ge n_w+m_w$, and that $\|\mL\|_{\fro}\le 2a\kappa$ and $\|\mR\|_{\fro}\le 2b\kappa$ by the inductive hypothesis,
\begin{align*}
    \|\mN\|_{\fro}&\le \|\mL\|_{\fro}+\|\mR\|_{\fro}+\sqrt{3n_w+3m_w}\\
    &\le 2a\kappa+2b\kappa+2\kappa\le 2(a+b+1)\kappa=2s\kappa
\end{align*}
\noindent
Similarly, we get:
\begin{align*}
    \|(\mN^{-1})_{I,J}\|_{\fro}&\le \|(\mL^{-1})_{I_L,J_L}\|_{\fro}+\|(\mR^{-1})_{I_R,J_R}\|_{\fro}\\
    &\le \kappa^a+\kappa^b\le 2\kappa^{a+b}\le \kappa^{a+b+1}=\kappa^s\\
    \|(\mN^{-1})_{I,}\|_{\fro}&\le \|(\mL^{-1})_{I_L,}\|_{\fro}+\|(\mR^{-1})_{I_R,}\|_{\fro}+\|(\mL^{-1})_{I_L,J_L}\|_{\fro}+\|(\mR^{-1})_{I_R,J_R}\|_{\fro}+\sqrt{n_w}\\
    &\le (5\kappa)^a+(5\kappa)^b+\kappa^a+\kappa^b+\kappa\le 5(5\kappa)^{a+b}\le (5\kappa)^{a+b+1}=(5\kappa)^s\\
    \|(\mN^{-1})_{,J}\|_{\fro}&\le \|(\mL^{-1})_{,J_L}\|_{\fro}+\|(\mR^{-1})_{,J_R}\|_{\fro}+\|(\mL^{-1})_{I_L,J_L}\|_{\fro}+\|(\mR^{-1})_{I_R,J_R}\|_{\fro}+\sqrt{m_w}\\
    &\le (5\kappa)^a+(5\kappa)^b+\kappa^a+\kappa^b+\kappa\le 5(5\kappa)^{a+b}\le (5\kappa)^{a+b+1}=(5\kappa)^s\\
    \|\mN^{-1}\|_{\fro}&\le \|\mL^{-1}\|_{\fro}+\|\mR^{-1}\|_{\fro}+\|(\mL^{-1})_{I_L,}\|_{\fro}+\|(\mL^{-1})_{,J_L}\|_{\fro}+\|(\mR^{-1})_{I_R,}\|_{\fro}+\|(\mR^{-1})_{,J_R}\|_{\fro}+\sqrt{n_w+m_w}\\
    &\le (10\kappa)^{2a+1}+(10\kappa)^{2b+1}+2(5\kappa)^a+2(5\kappa)^b+\kappa\le 7(10\kappa)^{2a+2b+1}\le (10\kappa)^{2a+2b+3}=(10\kappa)^{2s+1}
\end{align*}

\noindent Subtraction gates are the same as addition gates except that the $\mI_{,J_R}^{(n_R)}$ in the second row, third column of $\mN$ is replaced by $-\mI_{,J_R}^{(n_R)}$. The norm bound computations are then the same except for irrelevant sign changes.

\paragraph{Multiplication.}
Suppose the root gate is a multiplication gate.
Suppose the left child has $a\ge 1$ gates and the right child has $b\ge 1$ gates, where $s=a+b+1$.
Let $\mL$ be the $n_L\times n_L$ matrix and $\mR$ be the $n_R\times n_R$ matrix such that the outputs of the child gates are $(\mL^{-1})_{I_L,J_L}$ and $(\mR^{-1})_{I_R,J_R}$.

\begin{align*}
    \mN&=\begin{bmatrix}
        \mL&-\mI^{(n_L)}_{[n_L],J_L}\mI^{(n_R)}_{I_R,[n_R]}\\
        \mzero^{(n_R,n_L)}&\mR
    \end{bmatrix}\\
    \mN^{-1}&=\begin{bmatrix}
        \mL^{-1}&(\mL^{-1})_{,J_L}(\mR^{-1})_{I_R,}\\
        \mzero&\mR^{-1}
    \end{bmatrix}
\end{align*}

\noindent Selecting the rows of $\mN^{-1}$ with indices in $I$ is the same as taking the first row of blocks and left-multiplying by $\mI^{(n_L)}_{I_L,}$. Selecting columns with indices in $J$ is the same as taking the second column of blocks and right-multiplying by $\mI^{(n_R)}_{,J_R}$. Hence,
\begin{align*}
    (\mN^{-1})_{I,}&=\begin{bmatrix}
        (\mL^{-1})_{I_L,}&(\mL^{-1})_{I_L,J_L}(\mR^{-1})_{I_R,}
    \end{bmatrix}\\
    (\mN^{-1})_{,J}&=\begin{bmatrix}
        (\mL^{-1})_{,J_L}(\mR^{-1})_{I_R,J_R}\\
        (\mR^{-1})_{,J_R}
    \end{bmatrix}\\
    (\mN^{-1})_{I,J}&=\begin{bmatrix}
        (\mL^{-1})_{I_L,J_L}(\mR^{-1})_{I_R,J_R}\\
    \end{bmatrix}
\end{align*}

\noindent We again bound the Frobenius norms of these matrices using the bounds on $\|\mL\|_{\fro}$, $\|\mL^{-1}\|_{\fro}$, $\|(\mL^{-1})_{I_L,}\|_{\fro}$, $\|(\mL^{-1})_{,J_L}\|_{\fro}$, $\|(\mL^{-1})_{I_L,J_L}\|_{\fro}$, and similarly for $\mR^{-1}$, that we get from the inductive hypothesis.
The Frobenius norm of each block matrix is bounded by the sum of the Frobenius norms of its blocks. Using this together with the fact that $\|\mA\mB\|_{\fro}\le\|\mA\|_{\fro}\|\mB\|_{\fro}$, we get
\begin{align*}
    \|\mN\|_{\fro}&\le \|\mL\|_{\fro}+\|\mR\|_{\fro}+\sqrt{\min(n_L,n_R)}\\
    &\le 5a\kappa + 5b\kappa+\kappa\le 5(a+b+1)\kappa=5s\kappa\\
    \|(\mN^{-1})_{I,J}\|_{\fro}&\le \|(\mL^{-1})_{I_L,J_L}\|_{\fro}\|(\mR^{-1})_{I_R,J_R}\|_{\fro}\\
    &\le \kappa^a\kappa^b\le \kappa^{a+b+1}=\kappa^s\\
    \|(\mN^{-1})_{I,}\|_{\fro}&\le \|(\mL^{-1})_{I_L,}\|_{\fro}+\|(\mL^{-1})_{I_L,J_L}\|_{\fro}\|(\mR^{-1})_{I_R,}\|_{\fro}\\
    &\le (5\kappa)^a+\kappa^a(5\kappa)^b\le (5\kappa)^{a+b}+(5\kappa)^{a+b}\le (5\kappa)^{a+b+1}=(5\kappa)^s\\
    \|(\mN^{-1})_{,J}\|_{\fro}&\le \|(\mR^{-1})_{,J_R}\|_{\fro}+\|(\mR^{-1})_{I_R,J_R}\|_{\fro}\|(\mL^{-1})_{,J_L}\|_{\fro}\\
    &\le (5\kappa)^b+\kappa^b(5\kappa)^a\le (5\kappa)^{a+b}+(5\kappa)^{a+b}\le (5\kappa)^{a+b+1}=(5\kappa)^s\\
    \|\mN^{-1}\|_{\fro}&\le \|\mL^{-1}\|_{\fro}+\|\mR^{-1}\|_{\fro}+\|(\mL^{-1})_{,J_L}\|_{\fro}\|(\mR^{-1})_{I_R,}\|_{\fro}\\
    &\le (10\kappa)^{2a+1}+(10\kappa)^{2b+1}+(5\kappa)^a(5\kappa)^b\le 3(10\kappa)^{2a+2b+1}\le (10\kappa)^{2a+2b+3}=(10\kappa)^{2s+1}
\end{align*}
\end{proof}

\subsection{Proof of \Cref{thm:inverseds}}
\label{sec:main}

To obtain \Cref{thm:inverseds}, we will use the following data structures from previous work \cite{Sankowski04,BrandNS19}.
This previous work only considered finite fields or the real-RAM model, i.e., infinite precision with $O(1)$ time per arithmetic operation.
In \Cref{sec:inverse:proof}, we prove that these data structures also work with finite precision and $\tilde{O}(\log (\kappa/\epsilon))$-bit fixed-point arithmetic.

\begin{restatable*}{theorem}{dynamicInverse}\label{thm:inverseds}
    There exists several dynamic matrix inverse algorithms with the following operations.
    Each data structure initializes in $O(n^\omega \log \kappa/\epsilon)$ time on given accuracy parameters $\epsilon>0,\kappa > n$, and dynamic matrix $\mZ \in \R^{n\times n}$ that is promised to stay invertible throughout all updates with $\|\mZ\|_F,\|\mZ^{-1}\|_F\le \kappa$.

    The data structures have the following update/query operations
    \begin{enumerate}
        \item Support entry updates and entry queries in $O(n^{\eeEntryExp} \log(\kappa/\epsilon))$ time. \cite{BrandNS19} \label{item:fastupdate}
        \item Support entry updates in $O(n^{\neEntryUpdateExp} \log (\kappa/\epsilon))$ time and entry queries in $O(n^{\neEntryQueryExp} \log (\kappa/\epsilon))$ time. \cite{Sankowski04} \label{item:fastquery}
        \item Support rank-1 updates and return entire $\mZ^{-1}$ in $\tilde O(n^2 \log (\kappa/\epsilon))$ time. \cite{Sankowski04} \label{item:rank1}
        \item Support column updates and row queries in the offline model (the entire sequence of column indices and row queries is given at the start) in $\tilde O(n^{\omega-1} \log (\kappa/\epsilon))$ update and query time. \cite{BrandNS19} \label{item:offline}
    \end{enumerate}
    The outputs are all $\epsilon$-approximate, i.e., each entry is off by at most an additive $\pm\epsilon$.
\end{restatable*}

\begin{proof}[Proof of \Cref{thm:main}]
    Given the formula $f(\mm_1,...,\mm_s)$ and its input matrices $\mm_1,...,\mm_s$, where each $\mm_i$ is of size $n_i\times m_i$,
    let $n = \sum_s n_i + m_i$.

    \paragraph{Initialization}
    By \Cref{lemma:matrix-formula}, we can construct in $O(n^2)$ time a square $O(n)\times O(n)$ matrix $\mn$, where each $\mm_i$ is a subblock of $\mn$, and two sets $I,J\subset \Z$ with $(\mn^{-1})_{I,J} = f(\mm_1,...,\mm_s)$.

    The assumption of \Cref{thm:main} states that each $\|\mm_i\|_F\le\kappa$ and each result of an inversion gate within $f$ also has Frobenius norm bounded by $\kappa$.
    Thus, by \Cref{lemma:n-cond-bound}, we have $\log \|\mn\|_F$ and $\log \|\mn^{-1}\|_F$ bounded by $O(s \log \kappa)$.

    Depending on the type of update and query that we want (i.e., entry, column, row, etc),
    we run the respective data structure from \Cref{thm:inverseds} on $\mn$.

    In total, the initialization takes $\tilde O(n^\omega s \log (\kappa/\epsilon))$ time, dominated by the initialization of the data structure from \Cref{thm:inverseds}.

    \paragraph{Updates and Queries}
    Since each $\mm_i$ is a submatrix of $\mn$, entry, column, row, or rank-1 updates to any $\mm_i$ can be modelled by an entry, column, row, or rank-1 update to $\mn$.

    Likewise, querying an entry, a row, or column of $f(\mm_1,...,\mm_s)$ can be performed by querying an entry, or row, or column of $\mn^{-1}$, because submatrix $(\mn^{-1})_{I,J} = f(\mm_1,...,\mm_s)$.
    Further, the queries all have accuracy $\epsilon$ by \Cref{thm:inverseds}.

    Thus, the update and query complexity of \Cref{thm:main} is exactly as stated in \Cref{thm:inverseds}.
\end{proof}

\subsection{Example Application: Basic Solutions}
\label{subsec:basic_solution}

In this subsection, we give some example application of \Cref{thm:main}.
We show how these data structures could be used within iterative algorithms.
The example we provide is an algorithm by Beling and Megiddo \cite{Beling_Megiddo_1998} for converting some optimal solution of a linear program to an optimal \emph{basic} solution.
We start by defining basic solutions.

\begin{definition}[Basic Solution] 
    Given the standard form of linear constraints $\mA \mathbf{x} = \mathbf{b}, \mathbf{x}\geq \mathbf{0}$, where $\mA\in\mathbb{R}^{d\times n}$ is assumed to have linearly independent rows, $\mathbf{b}\in\mathbb{R}^d$, and $\mathbf{x}\in\mathbb{R}^n$, a solution $\mathbf{x}$ is said to be \emph{basic} if the set of columns $A_{:j}$ with $x_j\neq 0$ is linearly independent.
\end{definition}

\paragraph{Basis Crashing}
The \emph{basis crashing} problem seeks to find a basic solution to a system of linear constraints, given a non-basic solution. There is a naïve algorithm that solves the basis crashing problem in time $O(d^2 n)$, see \cite{Beling_Megiddo_1998}. 
This algorithm, similar to the simplex algorithm for linear programming,
repeatedly computes $\overline{\mA} = \mA_{,B}^{-1}\mA_{,k}$, where $\mA_{,B}$ are $d$ columns of $\mA$ and $k$ is some index (so $B\subset [n]$, $|B|=d$). The algorithm then checks if the resulting vector is entrywise non-negative.
In each iteration, one index is removed from $B$ and another is inserted, so we always have $|B|=d$. This is also called a pivoting step.
The previous $O(d^2 n)$ complexity stems from $O(d^2)$ per iteration over $O(n)$ iterations, where the analysis assumed real-RAM, i.e., infinite precision and $O(1)$ time per arithmetic operation. We show a $O(n^{2.528} \log(\kappa\max\det(A)))$ bound over word-ram, i.e., bit complexity, using approximate arithmetic, and finite bit precision. Here $\max\det(A)$, $\kappa$ are the maximum among all determinants and condition numbers of $d\times d$ submatrices of $\mA$. The maximum determinant is $\poly(n)$ for many linear programs modelling combinatorial problems.

With our data structure in  \Cref{application: general datastructures theorem}, we obtain tighter bounds on the time complexity and bit complexity of this algorithm.
For this, we use the formula $f(\ma,\ma,\ms) = (\mA \mS)^{-1} \mA$.
Here $\ms$ is a $n\times d$ matrix that is all 0, except for one $1$ per column, thus $\mA\mS$ is a $d\times d$ matrix consisting of $d$ columns of $\mA$. In other words, matrix $\ms$ selects column of $\mA$ to obtain $\mA_{,B}$.
Note that replacing an index in $B$ corresponds to changing $2$ entries in $\ms$. Further, getting $\mA_{,B}^{-1}\mA_{,k}$ for some $k$ requires us to query a column of $f(\ma,\ma,\ms)$.
With \Cref{application: general datastructures theorem} to maintain $f(\ma,\ma,\ms)$, this yields an improved runtime of $O(n^{1.528} \log \kappa/\epsilon)$ per iteration (to replace the two entries in $\mS$ and query the column) over $n$ iterations.
We are left with bounding $\epsilon$ such that we can guarantee the returned column vector to be correctly entry-wise positive (i.e., the error is small enough that we do not perform an incorrect decision).

Given $\mA\in\mathbb{R}^{d\times n}$, observe that a pivoting algorithm only needs to invert a smaller $d$-dimensional linear system. Next, recall that if the error $\epsilon$ of the linear system is at most $1/\det(\mA_{,B})$, then we can always round our matrix formula to the exact solution. Suppose matrix $\mA\in\mathbb{Z}^{d\times n}$ such that for all $(i,j)\in[d]\times[n]$, $|\mA_{ij}|<T$ for some $T\in\mathbb{R}$. Let $\lambda_i$ denote the $i$'th eigenvalue of $\mA_B$. It follows from our earlier entry-value bound on $\mA$ that $|\lambda_i|\leq mT$. Then, we have $\det(A_{,B}) = \prod_{i=1}^d \lambda_i \leq (dT)^d$. So, if the accuracy parameter $\epsilon < 1/(dT)^d$, we can always round our matrix formula to the exact solution. From \Cref{application: general datastructures theorem}, our data structure scales in $s \log \kappa/\epsilon$. Since the matrix formula is $(\mA\mS)^{-1}\mA$, we have immediately that the size of the formula is $s=O(1)$ and the Frobenius norm among all the inputs is at most $d\sqrt{T}$, while the Frobenius norm of $(\mA\mS)^{-1}$ can be bounded by $\kappa$ (the largest condition number of $d\times d$ submatrices of $\mA$).
So, the data structure scales in bit complexity $O(n^{\neEntryUpdateExp}\log (\kappa/\epsilon)) = O(n^{\neEntryUpdateExpPlusOne} (\log (\kappa) + d \log T))$.

We also get a general bound for the bit complexity of the basis crashing problem of $O(n^{\neEntryUpdateExpPlusOne}\log(\kappa\max\det(\mA)))$.
This is obtained by setting $\epsilon < 1/\max\det(\mA)$.

\section{Dynamic Determinant}
\label{sec:determinant}
In this section, our goal is to maintain $\det(f(\mm_1,\ldots,\mm_s))$ under the word-RAM model up to a multiplicative error of $1\pm\epsilon$ for $1>\epsilon>0$, when $\mm_1,\ldots,\mm_s$ go through low-rank updates. More specifically, the main theorem of this section is the following.

\thmDynDet*

To prove this theorem, we start by presenting some classical results about the determinant of matrices. \cref{lemma:det-schur-complement} characterize the determinant of block matrices and \cref{lemma:det-low-rank-update} characterize the change of determinant when a low-rank update is applied to a matrix.

\begin{lemma}
\label{lemma:det-schur-complement}
Let $\mm$ be a full-rank $n\times n$ matrix, $\mb$ and $\mc$ be $n\times m$ matrices, and $\md$ be an $m\times m$ matrics. Let
\[
\mm = \begin{bmatrix}
    \ma & \mb \\
    \mc^\top & \md
\end{bmatrix}.
\]
Then
\[
\det(\mm) = \det(\ma) \cdot\det(\md - \mc^\top \ma^{-1} \mb).
\]
\end{lemma}
\begin{proof}
Since $\ma$ is invertible, we can write a block LU decomposition of the matrix $\mm$.
\[
\mm = \begin{bmatrix}
\mi^{(n)} & \mzero \\
\mc^\top \ma^{-1} & \mi^{(m)}
\end{bmatrix}
\begin{bmatrix}
    \ma & \mzero \\
    \mzero & \md - \mc^\top \ma^{-1} \mb
\end{bmatrix}
\begin{bmatrix}
    \mi^{(n)} & \ma^{-1} \mb \\
    \mzero & \mi^{(m)}.
\end{bmatrix}
\]
Now note that by the multiplicativity of the determinant $\det(\mm)$ is just the product of determinants of the above three matrices. Finally since the determinant of the block triangular matrices with square matrices on the diagonal is the product of the determinants of the diagonal blocks, we have
\[
\det(\mm) = \det(\ma) \cdot \det(\md - \mc^\top \ma^{-1} \mb).
\]
\end{proof}

\noindent
Note that in the above $\md - \mc^\top \ma^{-1} \mb$ is just the Schur complement $\mm/\ma$ of the matrix $\mm$.

\begin{lemma}
\label{lemma:det-low-rank-update}
Let $\ma$ be a full-rank $n\times n$ matrix and $\matu$ and $\mv$ be $n\times m$ matrices such that $\ma + \matu \mv^\top$ is also full-rank. Then
\[
\det(\ma+\matu\mv^\top) = \det(\mi^{(m)}+ \mv^\top \ma^{-1} \matu) \cdot \det(\ma).
\]
\end{lemma}
\begin{proof}
Let 
\[
\mm = \begin{bmatrix}
\ma & -\matu \\
\mv^\top & \mi^{(m)}
\end{bmatrix}.
\]
By \cref{lemma:det-schur-complement}, since both $\ma$ and $\matu \mv^\top$ are invertible, we have
\[
\det(\mm) = \det(\ma)\cdot \det(\mi^{(m)} + \mv^\top \ma^{-1} \matu),
\]
and
\[
\det(\mm) = \det(\mi^{(m)})\det(\ma + \matu (\mi^{(m)})^{-1} \mv^\top).
\]
Then since $\det(\mi^{(m)})=1$ and $(\mi^{(m)})^{-1} = \mi^{(m)}$, we have
\[
\det(\ma+\matu\mv^\top) = \det(\mi^{(m)}+ \mv^\top \ma^{-1} \matu) \cdot \det(\ma).
\]
\end{proof}

\noindent
Equipped with these and by using \cref{lemma:matrix-formula}, we provide a simple formula for the determinant of a matrix formula for the determinant of a matrix formula.

\begin{lemma}
\label{lemma:det-matrix-formula}
Let $\mn$ be the matrix constructed for the matrix formula $f(\ma_1,\ldots,\ma_s)$ --- see \cref{lemma:matrix-formula} for construction of $\mn$. Moreover suppose $\mn$ is $n$-by-$n$.
Let $I$ and $J$ be the set of indices on the rows and columns of $\mn$ (with $\abs{I}=\abs{J}$ because otherwise determinant is not defined) such that
\[
f(\ma_1,\ldots,\ma_s) = 
\mi^{(n)}_{I,[n]} \mn^{-1} \mi^{(n)}_{[n],J}.
\]
Let 
\[
\widehat{\mn} = \begin{bmatrix}
    \mn & -\mi^{(n)}_{[n],J} \\
    \mi^{(n)}_{I,[n]} & \mzero
\end{bmatrix}.
\]
Then, if all inversions within $f(\ma_1,\ldots,\ma_s)$ exist, we have
\[
\det(f(\ma_1,\ldots,\ma_s)) = \frac{\det(\widehat{\mn})}{\det(\mn)}.
\]
\end{lemma}
\begin{proof}
By \cref{lemma:matrix-formula},
$\mn$ is invertible. Therefore by \cref{lemma:det-schur-complement},
\[
\det(\widehat{\mn}) = \det(\mn) \cdot \det(\mi^{(n)}_{I,[n]} \mn^{-1} \mi^{(n)}_{[n],J}) = \det(\mn) \cdot \det(f(\ma_1,\ldots,\ma_p)).
\]
Because $\mn$ is invertible, its determinant is nonzero and the result follows from the above equation.
\end{proof}

\Cref{lemma:det-matrix-formula} allows us to compute $\det(f(\mA_1,...,\mA_p))$ by computing both $\det(\mn)$ and $\det(\hat\mn)$. 
However, this fails when $\det(\hat\mn)=0$ as we would then divide by zero.
The following \Cref{lemma:det-of-mat-formula-matrix} shows that $\det(\hat\mn)$ is never 0, as long as $f(\mA_1,...,\mA_p)$ is well-defined, i.e., the formula does not attempt to invert a non-invertible matrix.

\begin{lemma}
\label{lemma:det-of-mat-formula-matrix}
Let $f(\ma_1,\ldots,\ma_s)$ be a matrix formula and $\mn$ be the matrix corresponding to it --- see \cref{lemma:matrix-formula}.
Suppose there are $r$ inversion gates in the formula and  $\mb_1,\ldots,\mb_r$ are the inputs to these inversion gates.
Then 
\[
\abs{\det(\mn)} = \prod_{i=1}^r \abs{\det(\mb_i)}.
\]
\end{lemma}
\begin{proof}
Note that the input gate for an $m \times n$ matrix $\mm$ constructs the following matrix.
\[
\mn = \begin{bmatrix}
\mi^{(m)} & \mm \\
\mzero & -\mi^{(n)}
\end{bmatrix}.
\]
The determinant of this matrix is $(-1)^n$. We now consider the inductive cases, i.e., inversion, product, and addition/subtraction of two matrices. For the product of two matrix formulas $f_{L}(\ma_1,\ldots,\ma_k)$ and $f_{R}(\ma_{k+1},\ldots,\ma_s)$, let $\ml$ and $\mr$ be the constructed matrices (which are square matrices by construction), respectively. Let $m$ and $n$ be the number of rows of $\ml$ and $\mr$, respectively. Let $I_L,J_L$ be the row and column indices corresponding to $f_{L}$ in $\ml^{-1}$, respectively, and similarly $I_R,J_R$ be the indices corresponding to $f_{R}$ in $\mr^{-1}$. Note that since the dimensions should be compatible, $\abs{J_L} = \abs{I_R}$. Then the constructed matrix for $f_{L}(\ma_1,\ldots,\ma_k) \cdot f_{R}(\ma_{k+1},\ldots,\ma_s)$ is 
\[
\mn = \begin{bmatrix}
\ml & -\mi^{(m)}_{[m],J_L} \mi^{(n)}_{I_R,[n]} \\
\mzero & \mr
\end{bmatrix}.
\]
Therefore by \cref{lemma:det-schur-complement}, $\det(\mn) = \det(\ml)\cdot \det(\mr)$.

We now discuss the addition/subtraction gates. Similar to the product gates, suppose $f_{L}(\ma_1,\ldots,\ma_k)$ and $f_{R}(\ma_{k+1},\ldots,\ma_s)$ are the formulas, $\ml$ and $\mr$ are the corresponding matrices, and $I_L,J_L,I_R,J_R$ are the corresponding indices. Because of the compatibility of the dimensions, we have $\abs{I_L}=\abs{I_R}$ and $\abs{J_L}=\abs{J_R}$. Let $p:=\abs{I_L}$ and $q:=\abs{J_L}$. Let $m$ and $n$ be the number of rows of $\ml$ and $\mr$, respectively. Then the constructed matrix for $f_{L}(\ma_1,\ldots,\ma_k) + f_{R}(\ma_{k+1},\ldots,\ma_s)$ is
\[
\mn = \begin{bmatrix}
\ml & \mzero & \mi^{(m)}_{[m],J_L} & \mzero \\
\mzero & \mr & \mi^{(n)}_{[n],J_R} & \mzero \\
\mzero & \mzero & -\mi^{(p)} & \mzero \\
\mi^{m}_{I_L,[m]} & \mi^{n}_{I_R,[n]} & \mzero & - \mi^{(q)}
\end{bmatrix}.
\]
Then by \cref{lemma:det-schur-complement}, we have
\begin{align}
\label{eq:det-sum-recursion}
\det(\mn) = \det(\ml)\det(\mr)\det(\ms),
\end{align}
where
\[
\ms = \mn/\begin{bmatrix}
    \ml & \mzero \\
    \mzero & \mr
\end{bmatrix}.
\]
By using, the block LU decomposition, one can check that the inverse of $\mn$ is as the following.
\[
\mn^{-1} = \begin{bmatrix}
\ml^{-1} & \mzero & (\ml^{-1})_{[m],J_L} & \mzero \\
\mzero & \mr^{-1} & (\mr^{-1})_{[n],J_R} & \mzero \\
\mzero & \mzero & \mi^{(p)} & \mzero \\
(\ml^{-1})_{I_L,[m]} & (\mr^{-1})_{I_R,[n]} & (\ml^{-1})_{I_L,J_L} + (\mr^{-1})_{I_R,J_R} & \mi^{(q)}
\end{bmatrix}.
\]
Therefore
\[
\ms^{-1} = \begin{bmatrix}
\mi^{(p)} & \mzero \\
(\ml^{-1})_{I_L,J_L} + (\mr^{-1})_{I_R,J_R} & \mi^{(q)}
\end{bmatrix}.
\]
Therefore $\det(\ms^{-1})=1$. Thus by \eqref{eq:det-sum-recursion}, $\det(\mn) = \det(\ml) \cdot \det(\mr)$.

We finally discuss the determinant for the inversion gates. Let $\mm$ be the matrix corresponding to the formula $f(\ma_1,\ldots,\ma_s)$, and $I$ and $J$ be the corresponding indices for $f$ in $\mm^{-1}$. By construction, $\mm$ is a square matrix. Suppose the number of rows of $\mm$ is $n$. Then the constructed matrix for $(f(\ma_1,\ldots,\ma_s))^{-1}$ is
\[
\mn = \begin{bmatrix}
\mm & \mi^{(n)}_{[n],J} \\
\mi^{(n)}_{I,[n]} & \mzero
\end{bmatrix}.
\]
By \cref{lemma:det-schur-complement},
\[
\det(\mn) = \det(\mm) \cdot \det((\mm^{-1})_{I,J}) = \det(\mm) \cdot \det(f(\ma_1,\ldots,\ma_s)).
\]
The result follows by noting the following. The determinant for an input gate (leaves of the formula tree) is one. The determinant for a product, addition, or subtraction gates is the product of the determinant of the children. The determinant of an inverse gate is the product of the determinant of its child and the determinant of its input.
\end{proof}

\noindent
The following is a direct result of \cref{lemma:det-of-mat-formula-matrix}.
\begin{remark}
If a matrix formula does not have an inversion, then the absolute value of the determinant of the constructed matrix is \emph{one}.
\end{remark}

\subsection{Stability of Computing the Determinant}
Our approach to computing the determinant of a static matrix $\ma$ is to compute a QR decomposition of $\ma$ and then take the product of diagonal entries of the matrix $\mr$, where $\ma = \mq \mr$. The following lemma gives an algorithm for QR decomposition under the word-RAM model.

\begin{lemma}[\cite{DemmelDH07}]
\label{lemma:qr}
Let $\ma\in\R^{n\times n}$ and $\epsilon>0$. There is an algorithm that computes matrices $\mr,\mw,\my \in \R^{n\times n}$ in $\Otil(n^{\omega}\log(1/\epsilon))$ time such that 
\[
(\mi - \mw \my + \delta\mq^\top)(\ma +\delta\ma) = \mr,
\]
where $\delta\mq$ and $\delta\ma$ are $n$-by-$n$ matrices with $\norm{\delta\mq}_{\fro} = O(\epsilon)$ and $\norm{\delta\ma}_{\fro} = O(\epsilon)\cdot\norm{\ma}_{\fro}$ and $\mr$ is an upper triangular matrix. Moreover setting $\mq:=(\mi - \mw \my + \delta\mq^\top)$, $\mq^\top \mq = \mi$.
\end{lemma}

\cref{lemma:qr} gives a backward stable QR decomposition. In other words, it computes a matrix $\mr$ such that $\det(\mr)=\det(\ma+\delta\ma)$ with $\norm{\delta\ma}_{\fro} = O(\epsilon)\cdot\norm{\ma}_{\fro}$. We also need the following lemma that bounds the product of the absolute values of eigenvalues with the product of singular values --- see \cite[p.~171]{horn_johnson_1991}.

\begin{lemma}[Weyl's inequality]
\label{lemma:weyl}
Let $\ma\in\R^{n\times n}$, $\sigma_1\geq \sigma_2 \geq \cdots \geq \sigma_n \geq 0$ be the singular values of $\ma$. Let $\lambda_1,\ldots,\lambda_n$ be the eigenvalues of $\ma$ such that $\abs{\lambda_1}\geq \abs{\lambda_2} \geq \cdots \geq \abs{\lambda_n} \geq 0$. Then for $k\in[n]$,
\[
\prod_{i=1}^k \abs{\lambda_i} \leq \prod_{i=1}^k \sigma_i.
\]
\end{lemma}

\noindent
The following lemma gives a bound on the determinant of a matrix that is purturbed by a small amount.

\begin{lemma}
\label{lemma:sum-det}
Let $\ma,\mx\in\R^{n\times n}$ and $n\geq 2$. Suppose $\ma$ is an invertible matrix. Let and $1\geq\epsilon\geq 0$ and  $ \hat{\epsilon}= \frac{\epsilon}{n^2\sigma_{\max}(\ma^{-1}\mx)}$. Then $\det(\ma+\hat{\epsilon}\mx)$ and $\det(\ma)$ have the same sign and 
\[
\abs{\det(\ma)} \cdot (1 + \hat{\epsilon} \cdot \tr(\ma^{-1} \mx) - \frac{\epsilon^2}{n})\leq \abs{\det(\ma+\hat{\epsilon}\mx)} \leq  \abs{\det(\ma)} \cdot (1 + \hat{\epsilon} \cdot \tr(\ma^{-1} \mx) + \frac{\epsilon^2}{n}).
\]
\end{lemma}
\begin{proof}
By multiplicativity of the determinant, we have
\begin{align}
\label{eq:sum-det-lemma-eq1}
\det(\ma + \hat{\epsilon} \mx) = 
\det(\ma) \det(\mi + \hat{\epsilon} \ma^{-1} \mx).
\end{align}
Let $\lambda_1,\ldots,\lambda_n$ with $\abs{\lambda_1}\geq \abs{\lambda_2} \geq \cdots \geq \abs{\lambda_n}$ be the eigenvalues of $\ma^{-1} \mx$. Then the eigenvalues of $\mi + \hat{\epsilon}\ma^{-1} \mx$ are $1+\hat{\epsilon}\lambda_1,\ldots,1+\hat{\epsilon}\lambda_n$. Therefore
\[
\det(\mi + \hat{\epsilon} \ma^{-1} \mx) = (1+\hat{\epsilon}\lambda_1)(1+\hat{\epsilon}\lambda_2) \cdots (1+\hat{\epsilon}\lambda_n).
\]
Then since $\tr(\ma^{-1} \mx) = \lambda_1+\ldots+\lambda_n$ and by \cref{lemma:weyl}, we have
\begin{align*}
\det(\mi + \hat{\epsilon} \ma^{-1} \mx) 
& \leq 
1 + \hat{\epsilon} \cdot \tr(\ma^{-1} \mx) + \sum_{i=2}^n \hat{\epsilon}^i \binom{n}{i} \abs{\lambda_1^i}
\\ & \leq
1 + \hat{\epsilon} \cdot \tr(\ma^{-1} \mx) + \sum_{i=2}^n \frac{\epsilon^i\cdot n^i \cdot \sigma_{\max}^i}{n^{2i} \sigma_{\max}^i}
\\ & = 
1 + \hat{\epsilon} \cdot \tr(\ma^{-1} \mx) + \sum_{i=2}^n \frac{\epsilon^i}{n^{i}}
 \\ & \leq 
1 + \hat{\epsilon} \cdot \tr(\ma^{-1} \mx) + \frac{\epsilon^2}{n},
\end{align*}
where $\sigma_{\max}:=\sigma_{\max}(\ma^{-1} \mx)$.
With a similar argument, one can see that
\[
1 + \hat{\epsilon} \cdot \tr(\ma^{-1} \mx) - \frac{\epsilon^2}{n} \leq \det(\mi + \hat{\epsilon} \ma^{-1} \mx)
\]
Note that by \cref{lemma:weyl}, $\tr(\ma^{-1} \mx) \geq -n \cdot  \sigma_{\max}$. Therefore 
\[
1 + \hat{\epsilon} \cdot \tr(\ma^{-1} \mx) - \frac{\epsilon^2}{n} \geq 0.
\]
Thus by \eqref{eq:sum-det-lemma-eq1}, the signs of $\det(\ma+\hat{\epsilon}\mx)$ and $\det(\ma)$ are the same. Moreover
\[
\abs{\det(\ma + \hat{\epsilon} \mx)} = 
\abs{\det(\ma)} \cdot \det(\mi + \hat{\epsilon} \ma^{-1} \mx)
\]
Then the result follows by applying the above bounds on $\det(\mi + \hat{\epsilon} \ma^{-1} \mx)$.
\end{proof}

Now we are equipped to bound the error of computing the determinant through QR decomposition under the word-RAM model for full-rank static matrices.

\begin{lemma}
\label{lemma:det-comp-by-qr}
Let $\ma\in\R^{n\times n}$ be full-rank and $0<\epsilon_1<1$. Then we can compute $d\in\R$ in $\Otil(n^{\omega} \log(\frac{\kappa(\ma)}{\epsilon_1}))$ time such that $d$ has the same sign as $\det(\ma)$ and
\[
(1-\epsilon_1) \cdot \abs{\det(\ma)} \leq \abs{d}\leq (1+\epsilon_1) \cdot \abs{\det(\ma)}.
\]
\end{lemma}
\begin{proof}
By \cref{lemma:qr}, we have $\det(\mr) = \det(\ma+\delta \ma)$. Suppose $\norm{\delta\ma}_{\fro}$ is so small that $n^2 \cdot \sigma_{\max}(\ma^{-1} \delta \ma)<1$. Then
\[
\epsilon := \hat{\epsilon}\cdot n^2 \cdot \sigma_{\max}(\frac{1}{\hat{\epsilon}} \cdot \ma^{-1} \delta\ma) = n^2 \cdot \sigma_{\max}(\ma^{-1} \delta\ma),
\]
is a valid $\epsilon$ for \cref{lemma:sum-det}. Then by \cref{lemma:sum-det}, we have
\begin{align*}
\abs{\det(\mr)} 
& = 
\abs{\det(\ma + \hat{\epsilon} (\frac{1}{\hat{\epsilon}}\cdot \delta\ma))} \leq  \abs{\det(\ma)} (1+ \hat{\epsilon} \tr(\frac{1}{\hat{\epsilon}}\ma^{-1}\delta\ma) + \frac{\epsilon^2}{n})
\\ & =
\abs{\det(\ma)} \cdot \left(1+ \tr(\ma^{-1}\delta\ma) + \frac{(n^2 \cdot \sigma_{\max}(\ma^{-1} \delta\ma))^2}{n} \right).
\end{align*}
Similarly by \cref{lemma:sum-det},
\begin{align*}
\abs{\det(\mr)} 
& \geq
\abs{\det(\ma)} \cdot  \left(1+ \tr(\ma^{-1}\delta\ma) - \frac{(n^2 \cdot \sigma_{\max}(\ma^{-1} \delta\ma))^2}{n} \right).
\end{align*}
Note that by the above construction and \cref{lemma:sum-det}, if  $n^2 \cdot \sigma_{\max}(\ma^{-1} \delta \ma)<1$, the sign of $\abs{\det(\mr)}$ is the same as the sign of $\abs{\det(\ma)}$. Now let
\[
\epsilon_2 = \frac{c\cdot \epsilon_1}{2n^2 \cdot\kappa(\ma)},
\]
where $c$ is the reciprocal of the constant corresponding to $\norm{\delta\ma}_{\fro} = O(\epsilon)\cdot\norm{\ma}_{\fro}$, i.e., $\norm{\delta\ma}_{\fro} \leq \frac{\epsilon_1}{2n^2 \cdot\kappa(\ma)} \norm{\ma}_{\fro}$.
By Cauchy-Schwarz inequality,
\[
\abs{\tr(\ma^{-1} \delta\ma)} \leq \norm{\ma^{-1}}_{\fro} \cdot  \norm{\delta\ma}_{\fro} \leq \frac{\epsilon_1}{2n \cdot \kappa(\ma)}  \cdot \norm{\ma^{-1}}_{2} \norm{\ma}_{2} \leq \frac{\epsilon_1}{2}.
\]
Moreover
\begin{align*}
\frac{(n^2 \cdot \sigma_{\max}(\ma^{-1} \delta\ma))^2}{n} 
& \leq 
n^3 \cdot \norm{\ma^{-1}}_2^2 \cdot \norm{\delta\ma}_2^2 
\leq 
n^3 \cdot \norm{\ma^{-1}}_2^2 \cdot \norm{\delta\ma}_{\fro}^2
\\ & \leq 
\frac{\epsilon_1^2}{2n^4 \kappa(\ma)^2} \cdot n^3 \cdot \norm{\ma^{-1}}_2^2 \cdot \norm{\ma}_{\fro}^2
\leq 
\frac{\epsilon_1^2}{4 n^4 \kappa(\ma)^2} \cdot n^4 \cdot \norm{\ma^{-1}}_2^2 \cdot \norm{\ma}_{2}^2
\\ & \leq \frac{\epsilon_1}{2}
\end{align*}
Therefore 
\[
1+ \tr(\ma^{-1}\delta\ma) + \frac{(n^2 \cdot \sigma_{\max}(\ma^{-1} \delta\ma))^2}{n} \leq 1+\epsilon_1,
\]
and 
\[
1+ \tr(\ma^{-1}\delta\ma) - \frac{(n^2 \cdot \sigma_{\max}(\ma^{-1} \delta\ma))^2}{n} \geq 1-\epsilon_1
\]
Combining the above, we have
\[
(1-\epsilon_1)\cdot \abs{\det(\ma)} \leq \abs{\det(\mr)} \leq (1+\epsilon_1)\cdot \abs{\det(\ma)}.
\]
Therefore we output $d=\det(\mr)$.
The running time consists of computing the QR factorization using \cref{lemma:qr} and then computing the determinant of $\mr$, which is the product of its diagonal entries. The dominating term is the former which by \ref{lemma:qr} and construction of $\epsilon$ is $\Otil(n^{\omega} \log(\frac{\kappa(\ma)}{\epsilon_1}))$.
\end{proof}

\subsection{Stability of Determinant Maintenance}

In this section, we analyze the stability and bit complexity of updating (via the matrix determinant lemma --- see \cref{lemma:det-low-rank-update}), the determinant of a matrix formula that goes through low-rank updates. Similar to maintaining the matrix formulas, the key is to show that certain bit complexity is sufficient to have a guarantee on the errors and that the accumulated error is not large. We need the following lemma to prove a general result regarding the maintenance of the determinant of matrix formulas.

\begin{lemma}[Forward-backward error connection \cite{MehrdadPV23}]
\label{lemma:forward-backward-equiv}
Let $\mm,\mn\in\R^{n\times n}$ be invertible matrices and $\kappa>1$ such that $\norm{\mn}_{\fro},\norm{\mn^{-1}}_{\fro} \leq \kappa$. Suppose $\norm{\mm-\mn}_{\fro} \leq \veps < \frac{1}{2\kappa}$. Then $\norm{\mm^{-1}}_{\fro} \leq 2 \kappa$, and
\[
\norm{\mm^{-1}-\mn^{-1}}_{\fro} \leq 2 \kappa^2 \cdot \veps.
\]
\end{lemma}

Equipped with this, we first prove the following general result and then combine it with \cref{application: general datastructures theorem} to prove \cref{thm:dynamic_det}.

\begin{theorem}\label{thm:determinant_stability}
    Let $f$ be a $s$-input matrix formula over $\R$ using $p$ gates that produces an invertible square matrix. Suppose matrix $\mN$, and index sets $I$, and $J$ are constructed as in \Cref{lemma:matrix-formula} so that $(\mN^{-1})_{I,J}=f(\mA_1,\ldots,\mA_s)$. Moreover let 
    \[
\widehat{\mn} = \begin{bmatrix}
    \mn & -\mi^{(n)}_{[n],J} \\
    \mi^{(n)}_{I,[n]} & \mzero
\end{bmatrix}.
\]
    Let the number of rows/columns of $\mn$ be $n$ and $\kappa\ge 2n$.
    Moreover, suppose  $\|\mA_1\|_{\fro},\ldots,\|\mA_s\|_{\fro}\le \kappa$, the Frobenius norms of outputs of the intermediary inverse gates are bounded by $\kappa$, and $\norm{f(\mA_1,\ldots,\mA_s)^{-1}}_{\fro} \leq \kappa$. 
    Let $\matu_1 \mv_1^\top,\matu_2 \mv_2^\top,\ldots,\matu_t \mv_t^\top$ be a series of $t$ low rank updates to $\mn$ corresponding to the low rank updates to the formula $f(\mA_1,\ldots,\mA_s)$ such that update $i\in[t]$ has rank $r_i$ and $\norm{\matu_1}_{\fro},\ldots,\norm{\matu_T}_{\fro},\norm{\mv_1}_{\fro},\ldots,\norm{\mv_t}_{\fro} \leq \kappa$. Moreover, suppose the above norm bounds hold after each update.
    
    Then for $0<\epsilon<1$ we can maintain $\det(f(\mA_1,\ldots,\mA_s))$ approximately to a multiplicative factor of $1\pm \epsilon$ over $t$ updates in time 
    \[
    \Otil\left(n^{\omega} s \cdot \log\left(\frac{\kappa\cdot t}{\epsilon}\right)+\sum_{i=1}^t T^{(1)}_i+T^{(2)}_i+r_i^{\omega} s \cdot \log\left(\frac{\kappa\cdot t}{\epsilon}\right)\right),
    \]
    where $T^{(1)}_i$ is the running time for \emph{implicitly} maintaining the matrices $\mn^{-1}$ and $\widehat{\mn}^{-1}$ at update $i$ to $\frac{\epsilon}{2t}$ accuracy (in terms of Frobenius norm error). By implicit maintenance, we mean that one can compute matrices $\mx_i$ and $\widehat{\mx}_i$ such that $\norm{\mx_i - (\mi^{(r_i)} + \mv_i^\top \mn^{-1} \matu_i)}_{\fro} \leq \frac{\epsilon}{2t}$ and $\norm{\widehat{\mx}_i - (\mi^{(r_i)} + \widehat{\mv}_i^\top \widehat{\mn}^{-1} \widehat{\matu}_i)}_{\fro} \leq \frac{\epsilon}{2t}$, where $\widehat{\mv}_i,\widehat{\matu}_i \in \R^{(n+|I|)\times r_i}$ are matrices obtained from $\mv_i$ and $\matu_i$ by concatenating $|I|$ zero rows. $T^{(2)}_i$ is the running time of computing $\mx_i$ and $\widehat{\mx}_i$ from the implicit representations.
\end{theorem}
\begin{proof}
By \cref{lemma:det-matrix-formula}, $\det(f(\ma_1,\ldots,\ma_s)) = \frac{\det(\widehat{\mn})}{\det(\mn)}$.
By assumptions and \cref{lemma:n-cond-bound},
$\norm{\mn}_{\fro} \leq \kappa^s$ and $\norm{\mn^{-1}}_{\fro} \leq (10\kappa)^{2s+1}$. Therefore since $\kappa\geq 2n$, $\widehat{\mn} \leq 2\kappa^s$. Moreover, by block inverse formula,
\[
\widehat{\mn}^{-1} = \begin{bmatrix}
\mn^{-1} - \mn^{-1}\mi^{(n)}_{[n],J} (\mi^{(n)}_{I,[n]} \mn^{-1} \mi^{(n)}_{[n],J})^{-1} \mi^{(n)}_{I,[n]}\mn^{-1}
& \mn^{-1} \mi^{(n)}_{[n],J} (\mi^{(n)}_{I,[n]} \mn^{-1} \mi^{(n)}_{[n],J})^{-1} 
\\
-(\mi^{(n)}_{I,[n]} \mn^{-1} \mi^{(n)}_{[n],J})^{-1} \mi^{(n)}_{I,[n]}\mn^{-1} & (\mi^{(n)}_{I,[n]} \mn^{-1} \mi^{(n)}_{[n],J})^{-1}
\end{bmatrix}.
\]
Note that $\mi^{(n)}_{I,[n]} \mn^{-1} \mi^{(n)}_{[n],J} = f(\mA_1,\ldots,\mA_s)$. Therefore by assumptions, triangle inequality, consistency of the norms and $\kappa\geq 2n$, we have
\[
\norm{\widehat{\mn}^{-1}}_{\fro} \leq 5\cdot (10\kappa)^{4s+4}.
\]
Note that a low-rank update updates the matrix $\mn$ with $\mn \leftarrow \mn +  \matu \mv^\top$, and it updates the matrix $\widehat{\mn}$ with $\widehat{\mn} \leftarrow \widehat{\mn} + \widehat{\matu} \widehat{\mv}^\top$ (we drop the subscript $i$ from $\matu_i,\mv_i,\widehat{\matu}_i,\widehat{\mv}_i$ for ease of notation). Then by \cref{lemma:det-low-rank-update}, 
\[
\det(\mn+\matu\mv^\top) = \det(\mn) \cdot \det(\mi^{(r)} + \mv^\top \mn^{-1} \matu)
\]
and
\[
\det(\widehat{\mn}+\widehat{\matu}\widehat{\mv}^\top) = \det(\widehat{\mn}) \cdot \det(\mi^{(r)} + \widehat{\mv}^\top \widehat{\mn}^{-1} \widehat{\matu}).
\]
Let 
\[
\mz = \begin{bmatrix}
    \mn & -\matu \\
    \mv^\top & \mi^{(r)}
\end{bmatrix}, ~~\text{and}~~ \widehat{\mz} = \begin{bmatrix}
    \widehat{\mn} & -\widehat{\matu} \\
    \widehat{\mv}^\top & \mi^{(r)}.
\end{bmatrix}
\]
Then by \cref{fact:block-matrix-inverse},
\[
\mz^{-1} = \begin{bmatrix}
(\mn +\matu \mv^\top)^{-1} & (\mn +\matu \mv^\top)^{-1} \matu \\
\mv^\top (\mn +\matu \mv^\top)^{-1} & \mi^{(r)} + \mv^\top (\mn +\matu \mv^\top)^{-1} \matu
\end{bmatrix}.
\]
Therefore by bounds on $\norm{\mn^{-1}}_{\fro}$ before and after updates $\norm{\mz^{-1}}_{\fro} \leq 5\cdot (10\kappa)^{2s+3}$ and similarly $\norm{\widehat{\mz}^{-1}}_{\fro} \leq 5\cdot (10\kappa)^{4s+6}$. Since $\mi^{(r)} + \mv^\top \mn^{-1} \matu$ is a Schur complement of $\mz$, 
\[
\norm{(\mi^{(r)} + \mv^\top \mn^{-1} \matu)^{-1}}_{\fro} \leq 5\cdot (10\kappa)^{2s+3}.
\] 
Similarly, we have
\[
\norm{(\mi^{(r)} + \widehat{\mv}^\top \widehat{\mn}^{-1} \widehat{\matu})^{-1}}_{\fro} \leq 5 \cdot (10\kappa)^{4s+6}.
\]
Moreover, note that by the norm bounds, we have
\[
\norm{\mi^{(r)} + \mv^\top \mn^{-1} \matu}_{\fro} \leq 2\cdot (10\kappa)^{2s+3}, ~~ \text{and} ~~ \norm{\mi^{(r)} + \widehat{\mv}^\top \widehat{\mn}^{-1} \widehat{\matu}}_{\fro} \leq 2 \cdot (10\kappa)^{4s+6}.
\]
By assumption, we can compute matrices $\mx$ and $\widehat{\mx}$ such that
\[
\norm{\mx - (\mi^{(r)} + \mv^\top \mn^{-1} \matu)}_{\fro} \leq \frac{\epsilon}{1000 \cdot t n^2 (10\kappa)^{2s+3}}, ~~ \text{and} ~~ \norm{\widehat{\mx} - (\mi^{(r)} + \widehat{\mv}^\top \widehat{\mn}^{-1} \widehat{\matu})}_{\fro} \leq \frac{\epsilon}{1000 \cdot t n^2 (10\kappa)^{4s+6}}.
\]
Then by triangle inequality and \cref{lemma:forward-backward-equiv},
\[
\norm{\mx}_{\fro}\leq 3\cdot(10\kappa)^{2s+3}, ~~ \text{and} ~~ \norm{\mx^{-1}}_{\fro} \leq 10\cdot (10\kappa)^{2s+3}.
\]
Similarly, we have
\[
\norm{\widehat{\mx}}_{\fro}\leq 3\cdot(10\kappa)^{4s+6}, ~~ \text{and} ~~ \norm{\widehat{\mx}^{-1}}_{\fro} \leq 10\cdot (10\kappa)^{4s+6}.
\]

Therefore by \cref{lemma:det-comp-by-qr}, for matrices $\mx$ and $\widehat{\mx}$, we can compute numbers $d_1$ and $d_2$ in $\Otil(r^{\omega}  \cdot s \cdot \log(\frac{\kappa\cdot t}{\epsilon}))$ such that the signs of $d_1$ and $\det(\mx)$ are the same and the signs of $d_2$ and $\det(\widehat{\mx})$ are the same, and moreover,
\begin{align}
\label{eq:det_x_bound}
\left(1-\frac{\epsilon}{200t}\right)\abs{\det(\mx)} \leq \abs{d_1} \leq \left(1+\frac{\epsilon}{200t}\right)\abs{\det(\mx)},
\end{align}
and
\[
\left(1-\frac{\epsilon}{200t}\right)\abs{\det(\widehat{\mx})} \leq \abs{d_2} \leq \left(1 + \frac{\epsilon}{200t}\right)\abs{\det(\widehat{\mx})}.
\]
Let 
\[
\widehat{\epsilon} = \frac{\epsilon / (100 t)}{n^2 \sigma_{\max}((\mi^{(r)} + \mv^{\top} \mn^{-1} \matu)^{-1}(\mx - (\mi^{(r)} +\mv^\top \mn^{-1} \matu)))}
\] 
and
$\delta \my = \frac{\mx - (\mi^{(r)} + \mv^\top \mn^{-1} \matu)}{\widehat{\epsilon}}$.
Then by \cref{lemma:sum-det}, the signs of $\det(\mi^{(r)} + \mv^{\top} \mn^{-1} \matu+\widehat{\epsilon}\delta\my)=\det(\mx)$ and $\det(\mi^{(r)} + \mv^{\top} \mn^{-1} \matu)$ are the same and
\[
\abs{\det(\mx)} \leq \abs{\det(\mi^{(r)} + \mv^{\top} \mn^{-1} \matu)} \cdot (1+\widehat{\epsilon}\cdot \tr((\mi^{(r)} + \mv^{\top} \mn^{-1} \matu)^{-1}\delta\my) + \frac{\epsilon^2}{10^4 t^2 n}).
\]
By Cauchy-Schwarz inequality, we have
\begin{align*}
\abs{\widehat{\epsilon}\cdot \tr((\mi^{(r)} + \mv^{\top} \mn^{-1} \matu)^{-1}\delta\my)} 
& = 
\abs{\tr((\mi^{(r)} + \mv^{\top} \mn^{-1} \matu)^{-1}(\mx - (\mi^{(r)} + \mv^\top \mn^{-1} \matu)))}
\\ & \leq 
\norm{(\mi^{(r)} + \mv^{\top} \mn^{-1} \matu)^{-1}}_{\fro} \cdot \norm{\mx - (\mi^{(r)} + \mv^\top \mn^{-1} \matu)}_{\fro}
\\ & \leq
\frac{\epsilon}{500 t n^2}.
\end{align*}
Therefore
\[
\abs{\det(\mx)} \leq \left(1+\frac{\epsilon}{250t}\right) \cdot \abs{\det(\mi^{(r)} + \mv^{\top} \mn^{-1} \matu)}.
\]
Combining this with \eqref{eq:det_x_bound}, we have
\[
\abs{d_1} \leq \left(1+\frac{\epsilon}{20 t}\right) \cdot \abs{\det(\mi^{(r)} + \mv^{\top} \mn^{-1} \matu)}.
\]
Similarly, one can show
\[
\left(1-\frac{\epsilon}{20 t}\right) \cdot \abs{\det(\mi^{(r)} + \mv^{\top} \mn^{-1} \matu)} \leq \abs{d_1}.
\]
Moreover since the signs of $d_1$ and $\det(\mx)$ are the same and the signs of $\det(\mx)$ and $\det(\mi^{(r)} + \mv^{\top} \mn^{-1} \matu)$ are the same, the signs of $d_1$ and $\det(\mi^{(r)} + \mv^{\top} \mn^{-1} \matu)$ are the same. Similarly by \cref{lemma:sum-det}, one can see that the signs of $d_2$ and $\det(\mi^{(r)} + \widehat{\mv}^{\top} \widehat{\mn}^{-1} \widehat{\matu})$ are the same and
\[
\left(1-\frac{\epsilon}{20t}\right)\abs{\det(\mi^{(r)} + \widehat{\mv}^\top \widehat{\mn}^{-1} \widehat{\matu})} \leq \abs{d_2} \leq \left(1+\frac{\epsilon}{20t}\right)\abs{\det(\mi^{(r)} + \widehat{\mv}^\top \widehat{\mn}^{-1} \widehat{\matu})}.
\]
Therefore
\[
\left(1-\frac{\epsilon}{6t}\right)\frac{\abs{\det(\mi^{(r)} + \mv^\top \mn^{-1} \matu)}}{\abs{\det(\mi^{(r)} + \widehat{\mv}^\top \widehat{\mn}^{-1} \widehat{\matu})}} \leq \abs{\frac{d_1}{d_2}} \leq \left(1+\frac{\epsilon}{6t}\right)\frac{\abs{\det(\mi^{(r)} + \mv^\top \mn^{-1} \matu)}}{\abs{\det(\mi^{(r)} + \widehat{\mv}^\top \widehat{\mn}^{-1} \widehat{\matu})}}
\]
Therefore, after $i$ updates, we have
\[
\exp\left(-\frac{i\cdot\epsilon}{3t}\right) \abs{\det(f(\ma_1,\ldots,\ma_s))} \leq \abs{d} \leq \exp\left(\frac{i\cdot\epsilon}{6t}\right) \abs{\det(f(\ma_1,\ldots,\ma_s))}
\]
Thus
\[
\left(1 -\frac{i \cdot \epsilon}{t}\right) \cdot  \abs{\det(f(\ma_1,\ldots,\ma_s))} \leq \abs{d} \leq \left(1 +\frac{i \cdot \epsilon}{t}\right) \cdot  \abs{\det(f(\ma_1,\ldots,\ma_s))}
\]
Since $i\leq t$ by assumption, the result follows.\qedhere

\end{proof}

\noindent
We are now equipped to prove our result for the determinant maintenance data structure.

\begin{proof}[Proof of \cref{thm:dynamic_det}]
First, note that all of the considered updates are of rank one. Therefore, all corresponding $r_i$'s in \cref{thm:determinant_stability} are equal to one. Moreover, after every $n$ update, we can restart the data structure by recomputing the inverses of matrices $\mn$ and $\widehat{\mn}$ in \cref{thm:determinant_stability} from scratch. This only poses an amortized $\Otil(n^{\omega-1} s \cdot \log(\kappa/\epsilon))$ per update and therefore, we can assume the number of updates $t$ is at most $n$. We now need to bound $T_i^{(1)}$ and $T_i^{(2)}$ (see \cref{thm:determinant_stability}) for the three different type of updates.

As discussed, our updates are of rank-1 and in the form $\vu_i \vecv_i^\top$. Entry updates correspond to the case where $\vu_i$ and $\vecv_i$ have exactly one nonzero entry. Therefore the updates $\mn \leftarrow \mn + \vu_i \vecv_i^\top$ and $\widehat{\mn} \leftarrow \widehat{\mn} + \widehat{\vu}_i \widehat{\vecv}_i^\top$ (see \cref{thm:determinant_stability} for definition of $\widehat{\vu}_i$ and $\widehat{\vecv}_i$), both correspond to entry updates. Moreover, in this case, computing $1+\vecv_i^\top \mn^{-1} \vu_i$ and $1+\widehat{\vecv}_i^\top \widehat{\mn}^{-1} \widehat{\vu}_i$ correspond to entry queries (note that we do not compute these exactly, see the proof of \cref{application: general datastructures theorem}). By \cref{application: general datastructures theorem}, this can be done in $\Otil(n^{\eeEntryExp} s \log \kappa/\epsilon)$ time per update.

In the case of column updates, $\vu_i$ is an arbitrary vector, but $\vecv_i$ has only one nonzero entry. In this case, the updates $\mn \leftarrow \mn + \vu_i \vecv_i^\top$ and $\widehat{\mn} \leftarrow \widehat{\mn} + \widehat{\vu}_i \widehat{\vecv}_i^\top$, both correspond to column updates and computing $1+\vecv_i^\top \mn^{-1} \vu_i$ and $1+\widehat{\vecv}_i^\top \widehat{\mn}^{-1} \widehat{\vu}_i$ correspond to row queries. By \cref{application: general datastructures theorem}, this can be done in $\Otil(n^{\neEntryUpdateExp} s \log \kappa/\epsilon)$ time per update. 

Finally, the general rank-1 updates correspond to rank-1 updates and returning all entries of matrices that are close (in terms of Frobenius norm) to $\mn^{-1}$ and $\widehat{\mn}^{-1}$, to compute $1+\vecv_i^\top \mn^{-1} \vu_i$ and $1+\widehat{\vecv}_i^\top \widehat{\mn}^{-1} \widehat{\vu}_i$ in time $\Otil(n^2 s \log \kappa/\epsilon)$. By \cref{application: general datastructures theorem}, these updates can be performed in $\Otil(n^{2} s \log \kappa/\epsilon)$ time per update. 
\end{proof}

\section{Dynamic Rank}
\label{sec:rank}
In this section, our goal is to maintain $\rank(f(\mM_1, \dots, \mM_p))$ for any matrix formula $f$ over a finite field $\F$ when the input matrices $\mM_1, \dots, \mM_p$ undergo low-rank updates. \cite{10.5555/1283383.1283397,BrandNS19} provides a dynamic algorithm for the dynamic maintenance of the rank of a dynamic matrix $\mm$ which supports entry updates in $O(n^{\eeEntryExp})$ time.
This section generalizes that result to general matrix formulas.

\thmDynRank*

\noindent \Cref{thm:dynamicRank} is obtained using a reduction from dynamic maintenance of the rank of a matrix formula to the dynamic maintenance of the determinant of the matrix formula in \Cref{lemma: det_g_rank_f} and \Cref{theorem: dynamic_rank_dynamic_determinant_reduction}. 
We then show how to maintain the determinant of any matrix formula $f$ over finite field $\F$ using the matrix determinant lemma and a result from \cite{Brand21}. Finally, in \Cref{application: maximum_matching_size}, we provide an application of the dynamic rank algorithm in the maintenance of the maximum matching size of a graph $\mathcal{G}$ that supports a number of updates. We first restate \cite[Lemma 4.1]{10.5555/1283383.1283397} for convenience below.

\begin{lemma}[{\cite[Lemma 4.1]{10.5555/1283383.1283397}}]\label{lemma: sankowski_lemma_4.1}
Let $\F$ be a finite field. Then, for any matrix $\overline{\mathbf{M}}$ of form \[\overline{\mathbf{M}} = \begin{bmatrix}
        \mathbf{M} & \mathbf{X} & \mathbf{0} \\
        \mathbf{Y} & \mathbf{0} & \mathbf{I}_n \\
        \mathbf{0} & \mathbf{I}_n & \mathbf{I}_k
    \end{bmatrix},\] where $\mathbf{M}\in\F^{n\times n}$, and $X$ and $Y$ are random matrices of size $n$ such that each entry is chosen uniformly at random from the extension field $\F^{ex}$ of $\F$, it must be true that
$\rank(\overline{\mathbf{M}})= 3n \iff \rank(\mathbf{M})\geq n-k$ with probability $1 - \frac{n}{|\F^{ex}|}$. Since $3n$ is the full-rank of $\overline{\mathbf{M}}$, this is equivalent to $\det(\overline{\mathbf{M}})\neq 0 \iff \rank(M)\geq n-k$ with probability $1 - \frac{n}{|\F^{ex}|}$.
\end{lemma}

\noindent We apply \Cref{lemma: sankowski_lemma_4.1} by ``embedding'' our matrix formula $f(\mM_1, \dots, \mM_p)$ into the form of matrix $\overline{\mM}$, where $\mM = f(\mM_1, \dots, \mM_p)$. This allows us to relate $\rank(f(\mM_1, \dots, \mM_p))$ to $\det(\overline{\mM})$. We formalize this notion in \Cref{lemma: det_g_rank_f}.

\begin{lemma}\label{lemma: det_g_rank_f}
Let $\F$ be a finite field and $\F^{ex}$ be a field extension of $\F$ with $\abs{\F^{ex}}\geq n$. Then, for any matrix formula $f(\mathbf{M}_1, \dots, \mathbf{M}_p) \in \F^{n\times n}$, consider the function 
\[
g(\mathbf{M}_1, \dots, \mathbf{M}_p, \mathbf{P}, \mathbf{Q}, \mathbf{R}_k) = \mathbf{P} f(\mathbf{M}_1, \dots, \mathbf{M}_p) \mathbf{Q} + \mathbf{R}_k,
\]
where $\mathbf{P}\in\R^{3n\times n}$ is such that $\mathbf{P}_{i,i}=1$ for $i\leq n$ and $0$ elsewhere, $\mathbf{Q}\in\R^{n\times 3n}$ is such that $\mathbf{Q}_{i,i}=1$ for $i\leq n$ and $0$ elsewhere, and $\mathbf{R}_k$ is a $3n$-by-$3n$ matrix given by:
\[\mathbf{R}_k = \begin{bmatrix}
    \mathbf{0} & \mathbf{X} & \mathbf{0} \\
    \mathbf{Y} & \mathbf{0} & \mathbf{I}_n \\
    \mathbf{0} & \mathbf{I}_n & \mathbf{I}_k
\end{bmatrix}\]
Here, $\mathbf{X}$ and $\mathbf{Y}$ are $n$-by-$n$ random matrices such that each entry is chosen uniformly at random from $\F^{ex}$, and for $k\in\{0,\dots,n\}$, $\mathbf{I}_k\in\R^{n\times n}$ is the partial identity matrix such that $(\mathbf{I}_k)_{i,i}=1$ for $i\leq k$ and $0$ elsewhere. 
Then, with probability $1 - \frac{n}{|\F^{ex}|}$, $\rank(f(\mathbf{M}_1, \dots, \mathbf{M}_p))\geq n-k$ if and only if $\det(g(\mathbf{M}_1, \dots, \mathbf{M}_p, \mathbf{P}, \mathbf{Q}, \mathbf{R}_k))\neq 0$.

\end{lemma}
\begin{proof}
Observe that 
\[g(\mathbf{M}_1, \dots, \mathbf{M}_p, \mathbf{P}, \mathbf{Q}, \mathbf{R}_k) = \begin{bmatrix}
    f(\mathbf{M}_1, \dots, \mathbf{M}_p) & \mathbf{X} & \mathbf{0} \\
    \mathbf{Y} & \mathbf{0} & \mathbf{I}_n \\
    \mathbf{0} & \mathbf{I}_n & \mathbf{I}_k
\end{bmatrix}.\]
Let $\mathbf{M}:=f(\mathbf{M}_1, \dots, \mathbf{M}_p)$ such that $g(\mathbf{M}_1,\dots,\mathbf{M}_p, \mathbf{P}, \mathbf{Q}, \mathbf{R}_k)$ is in the form of $\overline{\mathbf{M}}$ in \cref{lemma: sankowski_lemma_4.1}. Then, \cref{lemma: sankowski_lemma_4.1} yields the claim.\qedhere
\end{proof}

\noindent   Note that in order to guarantee small failure probability, we have to take $|\F^{ex}|\in\Omega(n^2)$.\\

\noindent In \Cref{theorem: dynamic_rank_dynamic_determinant_reduction}, we provide an algorithmic reduction from a dynamic algorithm, $\mathcal{A}$, that can maintain the determinant of any matrix formula over $\F$ to a dynamic algorithm, $\mathcal{B}$, that can maintain the rank of any matrix formula over $\F$, even when $\mathcal{A}$ is allowed to stop supporting new updates when the determinant of the matrix formula is $0$. 

\begin{theorem} \label{theorem: dynamic_rank_dynamic_determinant_reduction}
Given any dynamic determinant algorithm $\mathcal{A}$ that maintains $\det(g(\mathbf{M_1}, \dots, \mathbf{M}_p, \mathbf{P}, \mathbf{Q}, \mathbf{R}_k))$, where $g=\mathbf{P} f(\mathbf{M_1}, \dots, \mathbf{M}_p)\mathbf{Q}+\mathbf{R}_k$ for any formula $f$ in a finite field $\F$ while supporting either rank-1 updates, column updates, row updates, or entry updates to input matrices $\mathbf{M}_1, \dots, \mathbf{M}_p$ in worst-case time $\mathcal{U}$  while being allowed to stop supporting new updates if $\det(g(\mM_1,\dots,\mM_p,\mP,\mQ,\mR_k))=0$, there also exists a dynamic rank algorithm $\mathcal{B}$ that maintains $\rank(f(\mathbf{M}_1, \dots, \mathbf{M}_p))$ while supporting the same types of updates in worst-case time $\mathcal{U}$.
\end{theorem}
\begin{proof}
We want to maintain the invariant that $\rank(f(\mathbf{M}_1, \dots, \mathbf{M}_p)) = n-k$, with high probability, by using $\mathcal{A}$ to maintain $\det(g(\mathbf{M}_1, \dots, \mathbf{M}_p, \mathbf{P}, \mathbf{Q}, \mathbf{R}_k))$. For this, we present the following algorithm $\mathcal{B}$, where updates $u$ are handled via Algorithm \ref{alg:rank_maintenance}.

\RestyleAlgo{algoruled}
\IncMargin{0.15cm}
\begin{algorithm}
\label{alg:rank_maintenance}
\textbf{Variables:} $\mm_1,\ldots,\mm_p, \mP, \mQ, \mr_k, k$.

 \SetKwProg{updateProc}{procedure}{$(\mm_1,\ldots,\mm_p)$}{}
  \updateProc{\update}{
    add update $u$ using $\mathcal{A}$ \\
    \If{$\det(g(\mM_1, \dots, \mM_p, \mP, \mQ, \mR_k))=0$}{
    revert update $u$, \\
    \emph{(the rank of $f(\mathbf{M}_1, \dots, \mathbf{M}_p)$ decreases)},\\
    add the $(k+1)$'th $1$ to $\mathbf{I}_k$ in $\mathbf{R}_k$,\\
    set $k\leftarrow k+1$,\\
    add update $u$ using $\mathcal{A}$,\\
    }
    \ElseIf{$\det(g(\mM_1, \dots, \mM_p, \mP, \mQ, \mR_k))\neq 0$}{
    $(*)$ try to delete the $k$'th $1$ from $\mI_k$,\\
    \If{$\det(g(\mM_1, \dots, \mM_p, \mP, \mQ, \mR_k))\neq 0$}{
    \emph{(the rank of $f(\mathbf{M}_1, \dots, \mathbf{M}_p)$ increases)}, \\
    $k\leftarrow k-1$.
    }
    \Else{revert the update from $(*)$}}}
\caption{Rank Maintenance Algorithm}
\end{algorithm}

\noindent Note by \Cref{lemma: det_g_rank_f} that $\rank(f(\mathbf{M}_1, \dots, \mathbf{M}_p)) = n-k$ with probability $1 - \frac{n}{|\F^{ex}|}$ as we are maintaining $\det(g(\mathbf{M}_1, \dots, \mathbf{M}_p, \mathbf{P}, \mathbf{Q}, \mathbf{R}_k))\neq 0$. If $\det(g(\mM_1,\dots,\mM_p,\mP,\mQ,\mR_k))=0$, then line 16 of Algorithm \ref{alg:rank_maintenance} allows us to revert the update while maintaining the rank of $f(\mM_1,\dots,\mM_p)$. Thus, $\mathcal{B}$ can maintain $\rank(f(\mathbf{M}_1, \dots, \mathbf{M}_p))$ in worst-case time $\mathcal{U}$. \qedhere
\end{proof}

\thmDynRank*
\begin{proof}

From \Cref{lemma: det_g_rank_f} and \Cref{theorem: dynamic_rank_dynamic_determinant_reduction}, we know that we can maintain $\rank(f(\mM_1, \dots, \mM_p))$ to be $n-k$ for any matrix formula $f$ over finite field $\F$ with high probability, if we can maintain $\det(g(\mM_1, \dots, \mM_p, \mP, \mQ, \mR_k))\neq 0$, where $g(\mM_1, \dots, \mM_p, \mP, \mQ, \mR_k) = \mP f(\mM_1, \dots, \mM_p)\mQ + \mR_k$. From \Cref{lemma:det-matrix-formula}, we know that the determinant of any matrix formula is given by
\[\det(g(\mM_1, \dots, \mM_p, \mP, \mQ, \mR_k)) = \frac{\det(\widehat{\mathbf{N}})}{\det(\mN)}\]
Therefore, to maintain $\det(g(\mM_1, \dots, \mM_p, \mP, \mQ, \mR_k))$, it suffices to simultaneously maintain $\det(\widehat{\mN})$ and $\det(\mN)$ for entry-updates in $\F$.\\

\noindent  From the matrix determinant lemma (\cref{lemma:det-low-rank-update}), we have that for any invertible $n$-by-$n$ matrix $\mX$ and $n$-by-$m$ matrices $\mU$ and $\mV$, that $\det(\mX+\mU\mV^T) = \det(\mX)\det(\mI_m + \mV^T \mX^{-1}\mU)$. Similarly, for $n$-by-$1$ vectors $\mathbf{u}$ and $\mathbf{v}$, we have that $\det(\mX+\mathbf{u}\mathbf{v}^T) = \det(\mX)(1+\mathbf{v}^T \mX^{-1} \mathbf{u})$. Thus, for matrices $\widehat{\mN}$ and $\mN$ over finite field $\F$, we also have that for entry-updates $\mathbf{u}$ and $\mathbf{v}$ (of appropriate dimensions):
\[\det(\widehat{\mN}+\mathbf{u}\mathbf{v}^T)=\det(\widehat{\mN})(1+\mathbf{v}^T \widehat{\mN}^{-1} \mathbf{u})\]
\[\det(\mN+\mathbf{u}\mathbf{v}^T)=\det({\mN})(1+\mathbf{v}^T {\mN}^{-1} \mathbf{u})\]
Then, by letting $h(\mathbf{v},\widehat{\mN},\mathbf{u}) = 1+\mathbf{v}^T\widehat{\mN}^{-1}\mathbf{u}$ and $h'(\mathbf{v},\mN,\mathbf{u}) = 1+\mathbf{v}^T \mN^{-1}\mathbf{u}$ and maintaining $h$ and $h'$ over the finite field $\F$, we can maintain the determinant of $\widehat{\mN}$ and $\mN$, which is sufficient for maintaining the determinant of any matrix formula $f(\mM_1, \dots, \mM_p)$ over finite field $\F$. The theorem follows by noting that maintaining $h$ and $h'$ can be done in $O(n^{\eeEntryExp})$ update time via \cite[Corollary 4.2]{Brand21}.
\end{proof}

\subsection{Example Application: Dynamic Maximum Matching}
\label{subsec:dynamic_max_matching}

A notable application of \cref{theorem: dynamic_rank_dynamic_determinant_reduction} is in maintaining the maximum matching size of a graph $\mathcal{G}$, while supporting updates like inserting/removing edges, turning vertices on/off, and merging vertices. We formalize this application in \Cref{application: maximum_matching_size} using the random Tutte matrix of $\mathcal{G}$.

\begin{theorem} \label{application: maximum_matching_size}\label{thm:matching}
Given any dynamic rank algorithm that maintains $\rank(f(\mathbf{M_1}, \dots, \mathbf{M}_p))$ for any formula $f$ in a finite field $\F$ while supporting updates to input matrices $\mathbf{M}_1, \dots, \mathbf{M}_p$ in time $\mathcal{U}$, there also exists a dynamic maximum matching size algorithm for a graph $\mathcal{G}$ that supports updates in time $\mathcal{U}$ which inserts/removes an edge, turns vertices on/off, and merges vertices.
The dynamic algorithm is randomized and correct with high probability.
\end{theorem}

In particular, the size of a dynamic maximum matching can be maintained in $O(n^{\eeEntryExp})$ time per update.

\begin{proof}
    From \cite{Geelen_2000}, the number of vertices in the maximum matching of a graph $\mathcal{G}=(\mathcal{V},\mathcal{E})$, where $\mathcal{V}=[n]$, is equal to the rank of the random Tutte matrix $\mT$ of $\mathcal{G}$ over field $\F$ with high probability. The random Tutte matrix of $\mathcal{G}$ over field $\F$ is defined as:
    \[\mT_{i,j} = \begin{cases}x_{ij}, & (i,j)\in \mathcal{E} \text{ and }i<j \\
    -x_{ji}, & (i,j)\in \mathcal{E} \text{ and } i>j \\
    0, & \text{ otherwise}\end{cases}\]
Here, for each $i,j\in[n]$, $x_{ij}$ is a uniformly random element from the finite field $\F$. We then maintain the rank of the Tutte matrix with matrix formula $f(\mI_1, \mT, \mI_2) = \mI_1 \mT \mI_2$, where $\mI_1$ and $\mI_2$ are identity matrices of size $n$, with the following update procedures:

\begin{itemize}
    \item To insert an edge $(u,v)$ for $u,v\in \mathcal{V}$ (assuming $v>u$): set entry $\mT_{u,v}\leftarrow x_{u,v}$ and entry $\mT_{v,u}\leftarrow -x_{u,v}$,
    \item To remove edge $(u,v)$ for $u,v\in \mathcal{V}$: set entry $\mT_{u,v}\leftarrow 0$ and entry $\mT_{v,u}\leftarrow 0$,
    \item To turn on vertex $v\in \mathcal{V}$: set entry $(\mI_{1})_{v,v}\leftarrow 1$ and entry $(\mI_2)_{v,v}\leftarrow 1$,
    \item To turn off vertex $v\in \mathcal{V}$: set entry $(\mI_{1})_{v,v}\leftarrow 0$ and entry $(\mI_2)_{v,v}\leftarrow 0$,
    \item To merge vertices $u, v \in \mathcal{V}$: set entry $(\mI_1)_{u,v}\leftarrow 1$, entry $(\mI_2)_{v,u}\leftarrow 1$, entry $(\mI_1)_{v,v}\leftarrow 0$, and entry $(\mI_2)_{v,v}\leftarrow 0$.
\end{itemize}
The Tutte matrix of $\mathcal{G}$ over field $\F$ is maintained during these update procedures, since adding two rows (or two columns) to each other still yields a Tutte matrix as the sum of two i.i.d. elements from $\F$ is still an i.i.d. element from $\F$. Therefore, given these kinds of entry updates to the matrices $\mI_1, \mT, \mI_2$, our dynamic rank algorithm should maintain $\rank(f(\mI_1, \mT, \mI_2)) = \rank(\mI_1\mT\mI_2)$ with high probability in time $\mathcal{U}$, which in turn should maintain the size of the maximum matching of graph $\mathcal{G}$ with high probability in time $\mathcal{U}$. \qedhere
\end{proof}
\section{Conclusion}

We discussed the bit complexity and stability of maintaining arbitrary matrix formulas (with inversion, multiplication, and addition/subtraction) and their determinants. In addition, we provided data structures for maintaining the rank of matrices under finite fields and discussed a few applications for these. We believe our data structures and analysis would provide a useful and easy-to-use toolbox for designing iterative algorithms under the word-RAM model. For example, to extend optimization algorithms to the word-RAM model, one only requires to provide an analysis that what amount of errors in the steps can be tolerated without affecting the convergence. Some other applications are in computational geometry and computer algebra problems.

A compelling future direction is to analyze the bit complexity of more complex algorithms that use algebraic and matrix formulas and the required error bounds for these algorithms. One interesting example is the Gram-Schmidt walk introduced for discrepancy minimization \cite{bansal2018gram} that has many applications including experimental design \cite{harshaw2019balancing}. It is not clear how many bits are required to guarantee such a random walk constructs a good distribution.

Another compelling direction is to investigate whether our bit complexity bounds can be improved for certain problems. For example, in the basic solution application, we presented bounds that depend on both the maximum determinant and maximum condition number over all $d$-by-$d$ submatrices. We know that the maximum determinant is small for combinatorial problems and the maximum condition number is small for random matrices with high probability. Therefore it would be interesting to investigate whether one of these terms can be eliminated from the bit complexity bound.

\addcontentsline{toc}{section}{References}
\bibliographystyle{alpha}
\bibliography{main}

\newpage
\appendix

\section{Error Analysis of Dynamic Matrix Inverse}

\dynamicInverse

All these data structures already known and proven in previous work \cite{Sankowski04,BrandNS19}, but since they considered real-RAM or finite fields, they did not include an error analysis.
The data structures all rely on repeated application of the Woodbury-identity:
$$
(\mZ + \mU \mV^\top)^{-1} = \mZ^{-1} - \mZ^{-1} \mU (\mI + \mV^\top \mZ^{-1} \mU)^{-1} \mV^\top \mZ^{-1} .
$$
Note that if $\mZ^{-1}$ is only available approximately, then we only have approximate $\mD\approx\mI + \mV^\top \mZ^{-1} \mU$. Further, inverting $\mD$ introduces further approximation error.
Thus repeated application of this identity to maintain the inverse of $\mZ$ will keep increasing the approximation error of $\mZ^{-1}$.
To bound how much the approximation increases over time, we use the following \Cref{lem:woodbury_error} by \cite{MehrdadPV23}.

\begin{lemma}[{\cite{MehrdadPV23}}]\label{lem:woodbury_error}
    Let $\mZ,\omZ\in\R^{n\times n}$ be invertible matrices.
    Moreover let $\mU,\mV \in \R^{n\times m}$ such that $\mZ+\mU\mV^\top$ is invertible.
    Let $\kappa > n+m$ such that
    $$
    \|\mU\|_{\fro}, \|\mV\|_{\fro}, \|\mZ\|_{\fro}, \|\mZ^{-1}\|_{\fro}, \|\mZ+\mU\mV^\top\|_{\fro}, \|(\mZ+\mU\mV^\top)^{-1}\|_{\fro} \le \kappa
    $$
    and $0<\epsilon_D \le 1$, and $\|\omZ-\mZ\|_{\fro} \le 1$.
    If $\mD\in\R^{m\times m}$ is an invertible matrix such that $\|\mD^{-1} - (\mI + \mV^\top \omZ^{-1} \mU)^{-1}\|_{\fro} \le \epsilon_D$,
    then
    $$
    \|(\omZ^{-1} - \omZ^{-1}\mU\mD^{-1}\mV^\top\omZ^{-1})^{-1} - (\mZ + \mU\mV^\top)^{-1}\|_{\fro} \le 512 \kappa^{26}\epsilon_D+\|\omZ-\mZ\|_{\fro}.
    $$
\end{lemma}

We now use this lemma to bound how much our approximation increases when updating (an approximation of) the inverse via the Woodbury-identity:

\begin{corollary}[Error of rank-$m$ Woodbury]\label{cor:woodbury}
    Let $\mZ,\omZ\in\R^{n\times n}$ be invertible matrices.
    Moreover let $\mU,\mV \in \R^{n\times m}$ such that $\mZ+\mU\mV^\top$ is invertible.
    Let $\kappa > n+m$ such that
    $$
    \|\mU\|_{\fro}, \|\mV\|_{\fro}, \|\mZ\|_{\fro}, \|\mZ^{-1}\|_{\fro}, \|\mZ+\mU\mV^\top\|_{\fro}, \|(\mZ+\mU\mV^\top)^{-1}\|_{\fro} \le \kappa
    $$
    and $0<\epsilon \le 1$, and $\|\omZ-\mZ\|_{\fro} \le 1$.

    When explicitly given $\omZ^{-1},\mU,\mV$, then in time $O(\MM(n,m,n) \log \kappa/\epsilon)$ we can compute $\omZ'^{-1}$, where each entry of $\omZ'^{-1}$ uses at most $O(\log \kappa/\epsilon)$ bit and
    $$
    \|\omZ' - (\mZ + \mU\mV^\top)\|_{\fro} \le 513\kappa^{26}\epsilon+\|\omZ-\mZ\|_{\fro}.
    $$
\end{corollary}
\begin{proof}
    Given $\omZ^{-1},\mU,\mV$ we compute $(\mI + \mV^\top \omZ^{-1} \mU)^{-1}$ up to $O(\log \kappa/\epsilon)$-bit precision in $O(m^\omega \log \kappa/\epsilon)$ time, which yields a matrix $\mD$ with $\|\mD - (\mI + \mV^\top \omZ^{-1} \mU)^{-1}\|_{\fro} \le \epsilon$.
    We then compute $\omZ'^{-1}$ by calculating $\omZ^{-1} - \omZ^{-1} \mU \mD \mV^\top \omZ^{-1}$ in $O(\MM(n,m,n)\log \kappa/\epsilon)$ time, and rounding each entry to some $O(\log \kappa/\epsilon)$ bit.
    By \Cref{lem:woodbury_error} (and large enough $O(\log \kappa/\epsilon)$ bit-length) we have
    $$
    \|\omZ' - (\mZ + \mU\mV^\top)\|_{\fro} \le 513 \kappa^{26}\epsilon +\|\omZ-\mZ\|_{\fro}.
    $$
\end{proof}

\subsection{Proof of \Cref{thm:inverseds}}
\label{sec:inverse:proof}

Let $\mZ$ be the dynamic input matrix that changes over time. The data structure task is to maintain $\mZ^{-1}$.
The four data structures in \Cref{thm:inverseds} all work by repeated application of the Woodbury-identity.
Internally, they store some $\omZ^{-1}$ (possibly in implicit form), representing the current $\mZ^{-1}$. 
We use \Cref{lem:woodbury_error} and \Cref{cor:woodbury} to bound how much the error of $\omZ^{-1}$ grows over a sequence of updates.
This error is measured via $\|\omZ - \mZ\|_{\fro}$. However, note that the matrix $\omZ$ is never computed by the data structure. It only exists for analysis purposes, and it is actually $\omZ^{-1}$ that is stored by the data structures.

The proofs in this section show, by induction over the number of updates $k$, that $\|\omZ - \mZ\|_{\fro} \le O(k \epsilon' \poly(\kappa))$ for some $\epsilon' > 0$.
Since $\mZ$ is known (it is the input to the data structure), 
we can always restart the data structure after $n$ updates by computing $\mZ^{-1}$ (approximately) in $O(n^\omega \log \kappa/\epsilon)$ time.
This yields $k \le n \le \kappa$ and thus bounds the maximum error that might occur within the data structures.

The following \Cref{lem:transform} then implies that $\|\omZ^{-1} - \mZ^{-1}\|_{\fro} \le O(\epsilon'\poly(\kappa))$.
By picking some small enough $\epsilon' = O(\epsilon/\poly(\kappa))$, we can thus bound $\|\omZ^{-1} - \mZ^{-1}\|_{\fro} \le \epsilon$ throughout all updates.

\begin{lemma}[{\cite{MehrdadPV23}}]\label{lem:transform}
    Let $\mM,\mN\in\R^{n\times n}$ be invertible matrices and $\kappa>1$ such that $\|\mN\|_{\fro},\|\mN^{-1}\|_{\fro} \le \kappa$.
    Suppose $\|\mM-\mN\|_{\fro} \le \epsilon \le 1/(2\kappa)$.
    Then $\|\mM^{-1}\|_{\fro} \le 2\kappa$, and
    $$
    \|\mM^{-1} - \mN^{-1}\|_{\fro} \le 2\kappa^2\epsilon.
    $$
\end{lemma}

\subsubsection*{\Cref{item:rank1} of \Cref{thm:inverseds}, $O(n^2)$ update time.}
We now prove \Cref{item:rank1} of \Cref{thm:inverseds} (restated below as \Cref{lem:rank-1}). This algorithm explicitly maintains $\omZ^{-1}$ in memory, and updates it via the Woodbury identity. Thus \Cref{lem:rank-1}, follows from \Cref{cor:woodbury}, together with the error analysis we outlined at the start of \Cref{sec:inverse:proof}.

\begin{lemma}\label{lem:rank-1}
    There is a dynamic matrix inverse data structure that, given $\mZ\in\R^{n\times n}$ with $\|\mZ\|_{\fro},\|\mZ^{-1}\|_{\fro} \le \kappa$ and $\epsilon>0$, initializes in $O(n^\omega \log \kappa/\epsilon)$ time.
    The data structure support rank-1 updates to $\mZ^{-1}$ in $O(n^2 \log \kappa/\epsilon)$ time.
    After each update it returns a $\omZ^{-1}$ with $\|\omZ-\mZ\|_{\fro},\|\omZ^{-1}-\mZ^{-1}\|_{\fro}\le\epsilon$.
\end{lemma}

\begin{proof}[Proof of \Cref{lem:rank-1} (\Cref{item:rank1} of \Cref{thm:inverseds})]
    Let $\mZ$ be the input matrix and $\omZ^{-1}$ be the inverse matrix maintained by the data structure.
    We prove by induction over $k$, the number of updates, that
    $$
    \|\mZ-\omZ\|_{\fro} \le (k+1)\epsilon'.
    $$
    We reset the data structure every $n$ updates so that $k \le n$ and thus we have a hard bound $n\epsilon'$ on the error.
    For small enough $\epsilon' = \Theta(\poly(\epsilon/\kappa))$ and \Cref{lem:transform}, this then implies $\|\mZ^{-1} - \omZ^{-1}\|_{\fro} \le \epsilon$ after every update.

    \paragraph{Initialization}
    During initialization, we compute an approximate inverse $\omZ^{-1}$ with 
    $$
    \|\omZ - \mZ\|\le \epsilon'.
    $$ 
    in $O(n^\omega \log \kappa/\epsilon)$ time.
    
    \paragraph{Update}
    We receive two vectors $u,v$. By assumption on $\|\mZ\|_{\fro} \le \kappa$ throughout all updates, so $\|\mZ+uv^\top\|_{\fro} \le \kappa$ and thus $\|u\|_2,\|v\|_2 \le 2\kappa$.
    Let $\mZ'=\mZ+uv^\top$ the new matrix.
    We compute $\omZ'^{-1}$ via \Cref{cor:woodbury}, with
    $$\|\omZ' - \mZ'\|_{\fro} \le \epsilon' + \|\omZ-\mZ\|_{\fro}$$
    in $O(n^2 \log \kappa/\epsilon)$ time.

    \paragraph{Complexity}
    In addition to the $O(n^2 \log \kappa/\epsilon)$ time per update, we also pay $O(n^{\omega}\log \kappa/\epsilon)$ every $n$ updates for the restart. This adds an extra $O(n^{\omega-1} \log \kappa/\epsilon)$ amortized cost to the update time, and is subsumed by the quadratic cost.
    The amortized update time can be made worst-case via standard technique.
\end{proof}

\subsubsection*{\Cref{item:fastquery} of \Cref{thm:inverseds}, $O(n^{\neEntryUpdateExp})$ update time.}
While \Cref{lem:rank-1} (\Cref{item:rank1} of \Cref{thm:inverseds}) maintains $\omZ^{-1}$ explicitly in memory, the other data structure (e.g., \Cref{item:fastquery} of \Cref{thm:inverseds}) store this matrix only in implicit form.
Intuitively, \cite{Sankowski04} only performs a partial calculation of the Woodbury-identity. That is,
$$
(\mZ+\mU\mV^\top)^{-1} = \mZ^{-1} - \underbrace{\mZ^{-1} \mU}_{=:\mL} \underbrace{(\mI+\mV^\top\mZ^{-1}\mU)^{-1} \mV^\top \mZ^{-1}}_{=:\mR}
$$
The matrices $\mL$ and $\mR$ are computed and stored. After $k$ updates, we thus have the implicit representation of $\mZ'$ (where we for simplicity ignore the rounding errors for now)
$$
\mZ'^{-1} = \mZ^{-1} + \sum_{i=1}^k \mL^{(i)} \mR^{(i)}
$$
where $\mZ'$ is the current matrix and $\mZ$ is the matrix from $k$ updates ago, and $\mL^{(i)},\mR^{(i)}$ for $i=1,...,k$ are the matrices constructed for the $i\th$ update.

\begin{lemma}\label{lem:generalizedSankowski}
    There is a dynamic matrix inverse data structure that, given $\mZ\in\R^{n\times n}$ with $\|\mZ\|_{\fro},\|\mZ^{-1}\|_{\fro} \le \kappa$ and $\epsilon>0$ and $0\le\nu\le\mu\le1$, initializes in $O(n^\omega \log \kappa/\epsilon)$ time.
    The data structure support entry queries to $\mZ^{-1}$ in $O(n^{\mu} \log \kappa/\epsilon)$ query time.
    It also support updates that each change $n^\nu$ entries of $\mZ$ at once. The update time is $O((\MM(n,n^\mu,n^\nu) + \MM(n,n,n^\mu)/n^{\mu-\nu}))$.
    The error bounds are $\|\omZ-\mZ\|_{\fro},\|\omZ^{-1}-\mZ^{-1}\|_{\fro}\le\epsilon$ where $\omZ^{-1}$ are the values maintained by the data structure and $\mZ$ is the input matrix.
\end{lemma}

For $\nu=0,\mu=\neEntryQueryExp$ we obtain \Cref{item:fastquery} of \Cref{thm:inverseds}.
We need the more general \Cref{lem:generalizedSankowski} to later prove \Cref{item:fastupdate} of \Cref{thm:inverseds}

\begin{proof}
    Let $\mZ$ be the input matrix.
    The data structure maintains some $\omZ^{-1}$ approximating $\mZ^{-1}$.
    We prove by induction over $k$, the number of updates, that 
    \begin{align}
    \|\mZ - \omZ\|_{\fro} \le (k+1)\epsilon'. \label{eq:fastqueryerror}
    \end{align}
    We reset the data structure every $n$ updates so that $k \le n$ and thus we have a hard bound on the error.
    For small enough $\epsilon' = O(\poly(\epsilon/\kappa))$ and \Cref{lem:transform}, this then implies $\|\mZ^{-1} - \omZ^{-1}\|_{\fro} \le \epsilon$ after every update.

    \paragraph{Base Case}
    During initialization, we compute an approximate inverse $\omZ^{-1}$ with 
    $$
    \|\omZ - \mZ\|\le \epsilon'.
    $$ 
    in $O(n^\omega \log \kappa/\epsilon)$ time.
    Store this matrix as $\mM$.
    So \eqref{eq:fastqueryerror} is true for $k=0$.

    \paragraph{Induction Step}
    Assume \eqref{eq:fastqueryerror} holds for $k-1$ updates. We now argue it also holds after processing the $k\th$ update.
    
    We receive two matrices $\mU,\mV$ where each column as only one non-zero entry (since the update changes only $n^\nu$ entries of $\mZ$). 
    By assumption on $\|\mZ\|_{\fro} \le \kappa$ throughout all updates, so $\|\mZ+\mU\mV^\top\|_{\fro} \le \kappa$ and wlog $\|\mU\|_2,\|\mV\|_2 \le \kappa$.
    Let $\mZ'=\mZ+\mU\mV^\top$ be the new matrix.

    We compute matrices $\mL, \mR$ with $\mL = \omZ^{-1}\mU$, $\mR = \mD v^\top \omZ^{-1}$ where $\|\mD-(\mV^\top \omZ^{-1} \mU)^{-1}\|_{\fro} < \epsilon'/(512\kappa^{26})$.
    This takes $O(\MM(n,n^\nu,n^\mu) \log \kappa/\epsilon')$ time.

    By \Cref{lem:woodbury_error} we have $\|(\omZ^{-1} - \ell r)^{-1} - \mZ' \|_{\fro} < \epsilon' + \|\omZ - \mZ\|_{\fro} = (k+1)\epsilon'$.

    \paragraph{Time Complexity}
    Instead of computing $\omZ^{-1} - \mL \mR$ after each update, we store $\mL,\mR$ in memory.
    In particular, for $\mL^{(i)},\mR^{(i)}$ being the matrices constructed for the $i\th$ update, we have
    $$
    \| (\mM - \sum_{i=1}^k \mL^{(i)} \mR^{(i)})^{-1} - \mZ' \|_{\fro} < (k+1)\epsilon'
    $$
    Thus when our update computes some $\omZ^{-1}\mU$ and $\mV^\top\omZ^{-1}$, 
    it actually computes 
    $$
    (\mM - \sum_{i=1}^k \mL^{(i)} \mR^{(i)}) \mU ~~\text{ and }~~ \mV^\top(\mM - \sum_{i=1}^k \mL^{(i)} \mR^{(i)})
    $$
    By $\mU,\mV$ having only one non-zero entry per column and $n^\nu$ columns, this corresponds to computing $n^\nu$ rows and column of $\omZ^{-1}$.
    By stacking the matrices $\mL^{(i)}, \mR^{(i)}$ to form one large matrix, we can write
    $$
    \omZ^{-1} = (\mM - \sum_{i=1}^k \mL^{(i)} \mR^{(i)}) = \mM - [\mL^{(1)}|...|\mL^{(k)}] [\mR^{(1)}|...|\mR^{(k)}]^\top
    $$
    So computing the necessary rows and columns of $\omZ^{-1}$ takes $O(\MM(n,kn^\nu,n^\nu))$ arithmetic operations.
    We can assume that each operation takes $O(\log \kappa/\epsilon') = O(\log \kappa/\epsilon)$ time, by always rounding each $\mL^{(i)},\mR^{(i)}$ to some $O(\log \kappa/\epsilon')$ bit when storing them after each update. This increases the error by at most a constant factor.

    For the query complexity, note that calculating one entry of $\omZ^{-1}$ takes $O(kn^\nu)$ arithmetic operations.

    Every $k=n^{\mu-\nu}$ updates, we explicitly compute $\omZ^{-1}$ via a rank-$n^\mu$ update containing the past $n^{\mu-\nu}$ updates.
    This is done via \Cref{cor:woodbury} and takes $O(\MM(n,n^\mu,n) \log \kappa/\epsilon)$ time to obtain a new $\omZ^{-1}$ with $\|\omZ - \mZ\|_{\fro} \le (k+1)\epsilon'$.
    While this deletes all the $\mL^{(i)},\mR^{(i)}$, and thus bounds the query complexity bu $O(n^\mu \log \kappa/\epsilon)$,
    it does not reset the error bound, because we updated $\omZ^{-1}$ via the Woodbury-identity.

    To reset the error back to $\epsilon'$, every $n$ updates we perform a complete reset of the data structure, i.e.~compute $\omZ^{-1}$ from scratch in $O(n^\omega \log \kappa/\epsilon)$ time and set $k=0$.

    The amortized time complexity per update is thus
    $O((\MM(n,n^\mu,n^\nu) + \MM(n,n^\mu,n)/n^{\mu-\nu} + n^{\omega-1}) \log \kappa/\epsilon) = O((\MM(n,n^\mu,n^\nu) + \MM(n,n^\mu,n)/n^{\mu-\nu}) \log \kappa/\epsilon)$.
\end{proof}

\subsubsection*{\Cref{item:fastupdate} of \Cref{thm:inverseds}, $O(n^{\eeEntryExp})$ update time.}
The idea of \Cref{item:fastupdate} of \Cref{thm:inverseds} (\cite{BrandNS19}) is similar to before: We only partially execute the Woodbury identity, i.e.~we have
$$
(\mZ+\mU\mV^\top)^{-1} = \mZ^{-1} - \mZ^{-1} \mU \underbrace{(\mI+\mV^\top\mZ^{-1}\mU)^{-1}}_{\mC} \mV^\top \mZ^{-1}
$$
Here only the matrices $\mU,\mV,\mC$ is computed and stored. 
Further, consider if we start with $\mU=\mV$ both being a $0$-matrix with some $k$ columns.
Then, as we receive entry updates to $\mZ$, we instead update $\mU$ and $\mV$ by inserting one extra non-zero entry.
This corresponds to performing a rank-2 update to matrix $\mC$, so we can maintain $\mC$ via Woodbury-identity.
After $k$ updates, each column of $\mU,\mV$ has a non-zero entry and we reset the data structure, i.e.~recompute $\mZ^{-1}$ and set $\mU=\mV$ being all-0 matrices again.
For further efficiency increase, we can use \Cref{lem:generalizedSankowski} to update $\mZ^{-1}$ every $k$ updates instead of recomputing $\mZ^{-1}$. 

\begin{proof}[Proof of \Cref{item:fastupdate} of \Cref{thm:inverseds}]
    Let $\mZ$ be the input matrix.
    The data structure maintains some $\omZ^{-1}$ approximating $\mZ^{-1}$.

    The following invariant is maintained throughout all updates
    $$
    \omZ^{-1} =  \mM^{-1} - \mM^{-1}\mU \omC^{-1} \mV^\top \mM^{-1}
    $$
    where $\mM^{-1}$ is an approximation of $\mZ^{-1}$ at most $n^\nu$ updates ago, with $\|\mM - \mZ\|_{\fro} \le \epsilon'$.
    Further,
    \begin{align}
        \mC = \mI + \mV^\top \mM^{-1} \mU \label{eq:representation}
    \end{align}
    and we maintain a $\omC^{-1}$ approximating $\mC^{-1}$ with $\|\omC^{-1}-\mC^{-1}\|_{\fro} \le \epsilon'/(512\kappa^{26})$.
    Thus by accuracy of $\omC^{-1}$ and \Cref{lem:woodbury_error} we get $\|\omZ - \mZ\|_{\fro} \le \epsilon'$ and for small enough $\epsilon'$ \Cref{lem:transform} then implies $\|\omZ^{-1} - \mZ^{-1}\|_{\fro} < \epsilon$.
    
    \paragraph{Initialization}
    We get $\mM^{-1}$ by initializing \Cref{lem:generalizedSankowski} on $\mZ$ with accuracy $\epsilon'$.
    Further, we construct two all-0 matrices $\mU,\mV$ with $n^\nu$ columns each and let $\mC=\mI$ the $n^\nu\times n^\nu$ identity.
    We also run \Cref{lem:rank-1} on $\mC$.
    
    \paragraph{Update}
    Assume this is the $k\th$ update.
    We receive two vectors $u,v$ where each has only one non-zero entry (since the update changes one entry of $\mZ$). 
    By assumption on $\|\mZ\|_{\fro} \le \kappa$ throughout all updates, so $\|\mZ+uv^\top\|_{\fro} \le \kappa$ and $\|u\|_2,\|v\|_2 \le \kappa$.

    We update matrices $\mU,\mV$ by inserting $u,v$ into the $k\th$ column respectively.
    We update $\mC$ to maintain invariant \eqref{eq:representation}, i.e.~we pass two rank-1 updates into \Cref{lem:rank-1}
    to maintain $\omC^{-1}$ in $O(n^{2\nu} \log \kappa/\epsilon)$ time.
    Note that this rank-1 update requires us to read $O(n^\nu)$ entries of $\mM^{-1}$ by $u,v$ having only one non-zero entry.
    By \Cref{lem:generalizedSankowski} this takes $O(n^\nu \cdot n^{\mu} \log \kappa/\epsilon)$ time.

    After $n^\nu$ updates, we reset $\mU=\mV=0$ and $\mC=\mI$ and reinitialize \Cref{lem:rank-1} to maintain $\mC^{-1}$.
    Further, we update $\mM^{-1}$ by performing the past $n^\nu$ entry updates via \Cref{lem:generalizedSankowski} in $O((\MM(n,n^\mu,n^\nu) + \MM(n,n,n^\mu)/n^{\mu-\nu}) \log \kappa/\epsilon)$ time.
    
    In summary, our amortized time per update is
    $$
    O((
    n^{\mu+\nu}
    +
    \MM(n,n^\mu,n^\nu)
    +
    \MM(n,n,n^\mu)/n^\mu
    )\log \kappa/\epsilon
    ).
    $$
    This is $O(n^{\eeEntryExp}\log \kappa/\epsilon)$ for 
     $\nu=0.54294416$, $\mu=0.86118267$ and the rectangular matrix multiplication exponents by \cite{WilliamsXXZ23}.

    \paragraph{Query}
    To perform a query, by
    $$
    \omZ^{-1} =  \mM^{-1} - \mM^{-1}\mU \omC^{-1} \mV^\top \mM^{-1}
    $$
    we need to query $n^\nu$ entries of $\mM^{-1}$ to get the required row of $\mM^{-1}\mU$, and column of $\mV^\top \mM^{-1}$.
    This takes $O(n^{\mu+\nu}\log \kappa/\epsilon)$ by \Cref{lem:generalizedSankowski}.
    The extra product with $\omC^{-1}$ takes $O(n^{2\nu}\log \kappa/\epsilon)$ time and is subsumed by the previous cost.
\end{proof}

\subsubsection*{\Cref{item:offline} of \Cref{thm:inverseds}, $O(n^{\omega-1})$ update time.}
We now sketch the idea of \Cref{item:offline} in \Cref{thm:inverseds} assuming no rounding errors. 
The data structure is from \cite{BrandNS19}, so we here recap only how and why the data structure works, but we do not give a complete proof of correctness.
Afterward, we analyse the resulting approximation error when performing the calculations exactly.

Let $\mZ$ be the matrix for which we receive a sequence of column updates to $\mZ$ and row queries to $\mZ^{-1}$.
Let us assume that for this sequence we know ahead of time the column indices of the updates and row indices of the row queries, i.e.~we have some sequence $i_1,i_2,..., \in [n]$.

For $\ell=1,...,\log(n)$ we let $\mZ^{(\ell)}$ be a copy of $\mZ$ that is created every $2^\ell$ operations (ie.~updates to $\mZ$ or queries to $\mZ^{-1}$).
Thus we can write 
$$\mZ^{(\ell)} = \mZ^{(\ell-1)} + \mU^{(\ell)}(\mV^{(\ell)})^\top$$
where $\mU^{(\ell)},\mV^{(\ell)}$ are of size $n\times 2^{\ell}$. By our updates being column updates, $\mU^{(\ell)}$ is a (potentially) dense matrix and $\mV^{(\ell)}$ is a sparse matrix with at most one non-zero entry per column. In particular, we can assume this non-zero entry per column of $\mV$ is just $1$.

Now consider the Woodbury-identity and let us analyse which information about $(\mZ^{(\ell+1)})^{-1}$ we require to figure out some rows of $(\mZ^{(\ell)})^{-1}$.
$$
(\mZ^{(\ell)})^{-1} = (\mZ^{(\ell)})^{-1} - (\mZ^{(\ell)})^{-1} \mU^{(\ell)} (\mI+(\mV^{(\ell)})^\top(\mZ^{(\ell+1)})^{-1}\mU)^{-1} (\mV^{(\ell)})^\top (\mZ^{(\ell+1)})^{-1}
$$
Notice that to compute $(\mV^{(\ell)})^\top (\mZ^{(\ell+1)})^{-1}$, we just need $2^\ell$ rows of $(\mZ^{(\ell+1)})^{-1}$ by sparsity of $\mV^{(\ell)}$.
Further, we can compute 
$$
\mR^{(\ell)} := (\mI+(\mV^{(\ell)})^\top(\mZ^{(\ell+1)})^{-1}\mU)^{-1} (\mV^{(\ell)})^\top (\mZ^{(\ell+1)})^{-1}
$$ 
in $O(\MM(2^\ell,2^\ell,n))$ time.
Then by
$$
(\mZ^{(\ell)})^{-1} = (\mZ^{(\ell+1)})^{-1} - (\mZ^{(\ell+1)})^{-1} \mU^{(\ell)} \mR^{(\ell)}
$$
we can compute any set of $r$ rows of $(\mZ^{(\ell)})^{-1}$ in $O(\MM(r,2^\ell,n)$ time. 
Further, to compute these rows, we do not need to know the entire $(\mZ^{(\ell+1)})^{-1}$ but the rows of $(\mZ^{(\ell+1)})^{-1}$ that we attempt to compute from $(\mZ^{(\ell)})^{-1}$.

Note that, since we are given the sequence of column indices of updates (and row indices queries) ahead of time, we know which $2^{\ell-1}$ rows of $(Z^{(\ell)})^{-1}$ are required to compute $\mR^{(\ell-1)}$ in the future. 
And we know the $2^\ell$ indices of the future $2^\ell$ row queries.
Thus we can precompute all the necessary information of $(\mZ^{(\ell)})^{-1}$ in $O(\MM(2^\ell,2^\ell,n))$ operations right after computing $\mR^{(\ell)}$.
In particular, we can assume that the necessary rows of $(\mZ^{(\ell+1)})^{-1}$ to compute $\mR^{(\ell)}$ have been precomputed as well.

In total, we get an amortized time per update/query operation of
$$\underbrace{O\left(
\sum_{\ell=0}^{\log n}
\MM(2^\ell,2^\ell,n) / 2^\ell
\right)}_{\begin{array}{c}
\text{Compute $\mR^{(\ell)}$ and $2^\ell$ rows of $(\mZ^{(\ell})^{-1}$} \\
\text{every $2^\ell$ operations, for $\ell=0,...,\log n$}
\end{array}}
=
O\left(
n^{\omega-1}
\right)$$
Note that we never have $\ell > \log n$, because we can just compute $\mZ^{(\ell)}$ in $O(n^\omega)$ time once every $n$ operations.

We now proceed to bound the error, when the calculations are performed approximately.

\begin{proof}[Proof of \Cref{item:offline} in \Cref{thm:inverseds}]
    We argue
    \begin{align}
        \|\omZ^{(\ell)} - \mZ^{(\ell)}\|_{\fro} \le \epsilon' (1+\log(n)-\ell) \label{eq:offlineapprox}
    \end{align}
    Thus for $\epsilon' = \Theta(\poly(\epsilon^{-1},\kappa)/\log n)$ we get $\|(\omZ^{(0)})^{-1} - (\mZ^{(0)})\|_{\fro} \le \epsilon$ by \Cref{lem:transform}.
    Note that $\mZ^{(0)}$ is precisely matrix $\mZ$ after every update/query operation.

    We prove \eqref{eq:offlineapprox} by induction over $\ell$, starting with $\ell = \log n$ as base case.
    \paragraph{Base Case}
    We compute $(\omZ^{(\log n)})^{-1}$ once at the start, and then again once every $n$ operations.
    We can compute this matrix such that $\|\omZ^{(\log n)} - \mZ^{(\log n)}\|_{\fro} \le \epsilon'$ in $O(n^\omega \log \kappa/\epsilon$ time.

    \paragraph{Induction Step}
    Let $\ell$ be such that \eqref{eq:offlineapprox} holds for $\ell+1$, i.e.~we have
    $$
     \|\omZ^{(\ell+1)} - \mZ^{(\ell+1)}\|_{\fro} \le \epsilon' (1+\log(n) - \ell - 1).
    $$
    Now consider what happens when we create $\mZ^{(\ell)}$.
    We compute
    $$
    \mR^{(\ell)} := \mD (\mV^{(\ell)})^\top (\omZ^{(\ell+1)})^{-1}
    $$ 
    where $\mD$ is such that
    $$
    \|\mD - (\mI+(\mV^{(\ell)})^\top(\omZ^{(\ell+1)})^{-1}\mU)^{-1}\|_{\fro} \le \frac{1}{512 \kappa^{26}}.
    $$
    This $\mD$ can be computed in $O(\MM(2^\ell,2^\ell,n) \log \kappa/\epsilon)$ time, with the dominating cost being the calculation of $(\mV^{(\ell)})^\top (\omZ^{(\ell+1)})^{-1}$.

    By \Cref{lem:woodbury_error} we have
    $$
    \|((\omZ^{(\ell-1)})^{-1} -(\omZ^{(\ell-1)})^{-1} \mU^{(\ell)} \mR^{(\ell)})^{-1} - \mZ^{(\ell)}\|_{\fro} \le \epsilon' + \|\omZ^{(\ell)} - \mZ^{(\ell)}\|_{\fro} \le \epsilon' (1+\log(n) - \ell).
    $$
    Further, computing the $2^\ell$ rows of $(\mZ^{(\ell)})^{-1}$ that will be required in the future to answer queries and compute $\mR^{(\ell-1)}$, can be computed in $O(\MM(2^\ell,2^\ell,n)\log \kappa/\epsilon)$.

    The amortized cost per update is $O(n^{\omega-1} \log \kappa/\epsilon)$.
\end{proof}
\section{General Matrix Formula}

\begin{theorem}[Theorem 3.1 of \cite{Brand21}]
\label{lemma:matrix-formula}
    Given a matrix formula $f(\mA_1,...,\mA_p)$ over field $\mathbb{F}$, define $n:=\sum_{i\in V}n_i+m_i$ where $n_i\times m_i$ is the dimension of $\mm_i$.
    
    Then there exists a square matrix $\mN$ of size at most $O(n)\times O(n)$, where each $\mA_i$ is a block of $\mN$.
    Further, if $f(\mA_1,...,\mA_p)$ is well-defined (i.e., it does not attempt to invert a non-invertible matrix) then $(\mN^{-1})_{I,J}=f(\mA_1,\ldots,\mA_p)$.
    
    Constructing $\mN$ and sets $I,J$ from $f$ takes $O(n^2)$ time.
\end{theorem}

The construction of $\mN$ is given below. The construction is given by induction over the tree size, where the tree is representing the given formula. There is one case for each type of gate (i.e., node) that could be the root of the tree.

\paragraph{Input gate}
A tree of size one consists only of an input gate, representing the input matrix of the formula.
\begin{align*}
    \mN^{-1}=\mN&=\begin{bmatrix}
        \mI^{(n_i)}&\mA_i\\
        \mzero^{(m_i,n_i)}&-\mI^{(m_i)}
    \end{bmatrix}
\end{align*}
where $\mA_i$ is the input matrix.

\paragraph{Inverse gate}
Given an inverse gate, let $w$ be its child gate and $\mN'$ be the matrix constructed for the tree (formula) rooted at $w$, and let $I',J'$ be the indices specifying the submatrix of $\mN'^{-1}$ that represent the formula rooted at $w$.
\begin{align*}
    \mN&=\begin{bmatrix}
        \mN'&-\mI^{(n_{N'})}_{,J'}\\
        \mI^{(n_{N'})}_{I',}&\mzero^{(n_w,n_w)}
    \end{bmatrix}
\end{align*}

\begin{align*}
    \mN^{-1}&=\begin{bmatrix}
        \mN'^{-1}-(\mN'^{-1})_{,J'}(\mN'^{-1})_{I',J'}^{-1}(\mN'^{-1})_{I',}& (\mN'^{-1})_{,J'}(\mN'^{-1})_{I',J'}^{-1}\\
        -(\mN'^{-1})_{I',J'}^{-1}(\mN'^{-1})_{I',}&(\mN'^{-1})_{I',J'}^{-1}
    \end{bmatrix}
\end{align*}

\paragraph{Addition and Subtraction gates}
Given an addition gate $w$, let $\mL,\mR$ be the matrices constructed for the left and right subtree (formulas), and let $I_L,J_L,I_R,J_R$ be the indices specifying the submatrices $(\mL^{-1})_{I_L,J_L}$ and $(\mR^{-1})_{I_R,J_R}$ that represent the formula of the left and right child. Let $n_L\times n_L$ and $n_R \times n_R$ be the dimensions of $\mL,\mR$ respectively. Let $n_w\times m_w$ be the dimension of the matrices that the addition gate $w$ is supposed to add (i.e., $|I_R|=|I_L|=n_w$ and $|J_R|=|J_L|=m_w$).
    \begin{align*}
    \mN&=\begin{bmatrix}
        \mL&\mzero&\mI^{(n_L)}_{,J_L}&\mzero\\
        \mzero&\mR&\mI^{(n_R)}_{,J_R}&\mzero\\
        \mI^{(n_L)}_{I_L,}&\mI^{(n_R)}_{I_R,}&\mzero&\mI^{(n_w)}\\
        \mzero&\mzero&\mI^{(m_w)}&\mzero
    \end{bmatrix}\\
    \mN^{-1}&=\begin{bmatrix}
        \mL^{-1}&\mzero&\mzero&-(\mL^{-1})_{,J_L}\\
        \mzero&\mR^{-1}&\mzero&-(\mR^{-1})_{,J_R}\\
        \mzero&\mzero&\mzero&\mI^{(m_w)}\\
        -(\mL^{-1})_{I_L,}&-(\mR^{-1})_{I_R,}&\mI^{(n_w)}&(\mL^{-1})_{I_L,J_L}+(\mR^{-1})_{I_R,J_R}\\
    \end{bmatrix}
    \end{align*}

This is the same formula for $\mN$ as in \cite{Brand21} except the last two rows are swapped, then the last row and column are negated. This swaps the last two columns of $\mN^{-1}$, and then negates the last row and column.

The subtraction gate is the same as the addition gate but with $\mI^{(n_R)}_{,J_R}$ in the second row third column of $\mN$ replaced by $-\mI^{(n_R)}_{,J_R}$.

\paragraph{Multiplication gate}
Given a multiplication gate, let $\mL,\mR$ be the matrices constructed for the left and right subtree (formulas), and let $I_L,J_L,I_R,J_R$ be the indices specifying the submatrices $(\mL^{-1})_{I_L,J_L}$ and $(\mR^{-1})_{I_R,J_R}$ that represent the formula of the left and right child. Let $n_L\times n_L$ and $n_R \times n_R$ be the dimensions of $\mL,\mR$ respectively. 
\begin{align*}
    \mN&=\begin{bmatrix}
        \mL&-\mI^{(n_L)}_{[n_L],J_L}\mI^{(n_R)}_{I_R,[n_R]}\\
        \mzero^{(n_R,n_L)}&\mR
    \end{bmatrix}\\
    \mN^{-1}&=\begin{bmatrix}
        \mL^{-1}&(\mL^{-1})_{,J_L}(\mR^{-1})_{I_R,}\\
        \mzero&\mR^{-1}
    \end{bmatrix}
\end{align*}

\end{document}